\documentclass{emulateapj}
\usepackage{apjfonts}

\newcommand{\dfn}{D$_{n}(4000)$}
\newcommand{\ha}{H$\alpha~$}
\newcommand{\hb}{H$\beta~$}

\newcommand{\airx}{$\widehat{A}_{IRX}$}

\begin{document}
\title{Ultraviolet, Optical, and Infrared Constraints on Models of Stellar Populations and Dust Attenuation}


\author{
Benjamin D. Johnson\altaffilmark{1}
David Schiminovich\altaffilmark{1},
Mark Seibert\altaffilmark{2},
Marie Treyer\altaffilmark{3,4},
D. Christopher Martin\altaffilmark{4},
Tom A. Barlow\altaffilmark{4}, 
Karl Forster\altaffilmark{4},
Peter G. Friedman\altaffilmark{4},
Patrick Morrissey\altaffilmark{4},
Susan G. Neff\altaffilmark{5},
Todd Small\altaffilmark{4},
Ted K. Wyder\altaffilmark{4},
Luciana Bianchi\altaffilmark{6},
Jose Donas\altaffilmark{3},
Timothy M. Heckman\altaffilmark{7},
Young-Wook Lee\altaffilmark{8},
Barry F. Madore\altaffilmark{2},
Bruno Milliard\altaffilmark{3},
R. Michael Rich\altaffilmark{9},
Alex S. Szalay\altaffilmark{7},
Barry Y. Welsh\altaffilmark{10}, 
Sukyoung K. Yi \altaffilmark{8},
}

\altaffiltext{1}{Department of Astronomy, Columbia University, New York, NY 10027}
\altaffiltext{2}{Observatories of the Carnegie Institution of Washington, 813 Santa Barbara St., Pasadena, CA 91101}
\altaffiltext{3}{Laboratoire d'Astrophysique de Marseille, BP 8, Traverse du Siphon, 13376 Marseille Cedex 12, France}
\altaffiltext{4}{California Institute of Technology, MC 405-47, 1200 East California Boulevard, Pasadena, CA 91125}
\altaffiltext{5}{Laboratory for Astronomy and Solar Physics, NASA Goddard Space Flight Center, Greenbelt, MD 20771}
\altaffiltext{6}{Center for Astrophysical Sciences, The Johns Hopkins University, 3400 N. Charles St., Baltimore, MD 21218}
\altaffiltext{7}{Department of Physics and Astronomy, The Johns Hopkins University, Homewood Campus, Baltimore, MD 21218}
\altaffiltext{8}{Center for Space Astrophysics, Yonsei University, Seoul 120-749, Korea}
\altaffiltext{9}{Department of Physics and Astronomy, University of California, Los Angeles, CA 90095}
\altaffiltext{10}{Space Sciences Laboratory, University of California at Berkeley, 601 Campbell Hall, Berkeley, CA 94720}

\begin{abstract}
The color of galaxies is a fundamental property, easily measured, that constrains models of galaxies and their evolution. Dust attenuation and star formation history (SFH) are the dominant factors affecting the color of galaxies.  Here we explore the empirical relation between SFH, attenuation, and color for a wide range of galaxies, including early types.  These galaxies have been observed by \emph{GALEX}, SDSS, and \emph{Spitzer}, allowing the construction of measures of dust attenuation from the ratio of infrared (IR) to ultraviolet(UV) flux and measures of SFH from the strength of the 4000\AA~break.  The empirical relation between these three quantities is compared to models that separately predict the effects of dust and SFH on color.  This comparison demonstrates the quantitative consistency of these simple models with the data and hints at the power of multiwavelength data for constraining these models. The UV color is a strong constraint; we find that a Milky Way extinction curve is disfavored, and that the UV emission of galaxies with large 4000\AA~break strengths is likely to arise from evolved populations.  We perform fits to the relation between SFH, attenuation, and color.  This relation links the production of starlight and its absorption by dust to the subsequent reemission of the absorbed light in the IR.  Galaxy models that self-consistently treat dust absorption and emission as well as stellar populations will need to reproduce these fitted relations in the low-redshift universe.

\end{abstract}


\section{Introduction}
Since the work of \citet{tinsley68} the location of galaxies in color-color and color-magnitude diagrams has proven to be an important indicator of the stellar content of galaxies, just as the location of stars in these same diagrams indicates their stellar spectral type.  This stellar content is in turn used to constrain the star formation histories (SFHs) of galaxies, which is an important step toward understanding the diversity of galaxies observed in the universe today.  Color-color diagrams are still used at high redshift (and low), where in addition to the stellar populations and dust attenuation, the redshift is estimated \citep{ss92, mobasher_photoz, wuyts06}.  

To obtain strong constraints on the constituent stellar populations with color-color diagrams, it is important to know the properties of the dust attenuation in galaxies. Selective attenuation can cause a large amount of reddening that can mask the differences in color due to differences in stellar populations.  Such attenuation is often treated as a reddening `vector' in color-color space \citep[e.g.,][]{kauffmann03a}.  The direction of the vector is given by the shape of the attenuation curve (often assumed to be that of \citet{calzetti94}), while the length of the vector is inferred indirectly.  Without independent knowledge of the length of the reddening vector, the power of color-color diagrams to probe the constituent stellar populations of galaxies is reduced.

The proper treatment of dust attenuation is of paramount importance when considering the constraints imposed by the color-magnitude diagram (CMD) on theories of galaxy formation \citep[e.g.,][]{noeske07b,labbe07}.  This is especially true since there is a strong correlation of attenuation with luminosity \citep{WH96} and stellar mass \citep[e.g.,][]{salim07}.  These correlations can induce or alter trends between the derived SFH and stellar mass of galaxies.  Similarly, dust attenuation also affects comparisons between the predictions of semi-analytic modelling and observations of galaxies at any wavelength.  

With the \emph{Galaxy Evolution Explorer} (\emph{GALEX}) and \emph{Spitzer Space Telescope} it is now possible to obtain ultraviolet (UV) and infrared (IR) fluxes for large samples of galaxies.  The combination of the UV and IR flux constrains the amount of dust attenuation in galaxies (i.e. the length of the reddening vector). The UV is due to emission from young stars that is not attenuated by dust, while the IR measures the emission from young stars that has been absorbed by dust and reradiated.  With this independent knowledge of the dust attenuation, which significantly affects the observed color of galaxies, these colors can be used to more sensitively probe the stellar content of galaxies and test models of galaxy formation.  New models of galaxy spectra are being developed and improved that treat dust attenuation, dust emission, and stellar populations in a consistent manner, building from the work of \citet{charlot00}.  These UV through IR models \citep[e.g.,][]{grasil, sunrise} promise to improve the treatment of attenuation and sharpen the constraints imposed by UV and IR observations on theories and semi-analytic models of galaxy formation.

In \citet{dissecting} we presented the relation between broadband UV through near-IR colors, dust attenuation (as measured by the infrared excess, IRX), and SFH (as measured by the 4000\AA~break) for a sample of galaxies observed with \emph{GALEX}, the Sloan Digital Sky Survey (SDSS), and \emph{Spitzer}. Here we present this same relation, with deeper UV data, for a larger sample of galaxies and a greater variety of broadband colors. We show that this relation serves as a useful diagnostic of galaxy properties, and an empirical constraint on models of dust attenuation and star formation (SF). We compare our results to models of galaxy spectra \citep{BC03} and dust attenuation \citep{WG00}. In addition, we consider where dwarf galaxies and IR- and UV-luminous galaxies fall in this diagram. This comparison of special classes of galaxies to a more representative sample places them in context, and aids the interpretation of the relation.

We present parameterized fits to this relation between color, the 4000\AA~break strength, and IRX. These fits may be used to determine the dust attenuation in galaxies where only optical spectra are available, even in the absence of measurable \ha~and \hb~lines.  In addition, these relations provide an empirical guide for self-consistent models of the UV through IR spectra of galaxies.

\section{Data}
Our primary sample is selected from the SDSS main galaxy spectroscopic sample \citep{strauss02}.  These galaxies have been imaged in five optical bands ($ugriz$) and observed spectroscopically from 3800\AA~to 9000\AA~(observed frame) in a 3\arcsec diameter aperture as part of the SDSS.  

In this paper we consider a subset of the SDSS spectroscopic galaxy sample that has been observed by both \emph{GALEX} and \emph{Spitzer}.  These galaxies are located in two different contiguous regions, the Lockman Hole and the \emph{Spitzer} extragalactic First Look Survey (FLS).  Details of the observations for each of these fields are given in Table \ref{tbl:obs}.  The \emph{Spitzer} observations include four-band Infrared Array Camera \citep[IRAC;][]{irac} imaging (3.6\micron, 4.5\micron, 5.8\micron, and 8\micron) as well as 24 and 70 \micron~Multiband Imaging Photometer for Spitzer \citep[MIPS;][]{mips} imaging. The \emph{GALEX} observations include imaging at 1517\AA~(far-UV, $f$) and 2267\AA~(near-UV, $n$). All quoted magnitudes are on the AB system and have been corrected for foreground extinction (which is especially low in the case of the Lockman Hole) according to the maps of \citet{SFD}.


\begin{deluxetable}{cccccc}
\tablecolumns{6}
\tablecaption{Observations \label{tbl:obs}
}
\tablehead{
\colhead{Field Name} & \colhead{Size}  & \colhead{N$_{obs}$\tablenotemark{a}} & \colhead{N$_{det}$\tablenotemark{b}} & \colhead{N$_{det}$\tablenotemark{b}} & \colhead{N$_{smpl}$\tablenotemark{c}} \\
\colhead{}   & \colhead{(deg$^{2}$)} & \colhead{} & \colhead{$f,n<25$} & \colhead{$m_{24}<19.5$} &  \colhead{}
}
\startdata
Lockman Hole & $\sim 9$ & 872 & 792 & 819 & 721  \\
FLS & $\sim 3$ & 186 & 147 & 158 & 118 \\
\enddata
\tablecomments{The median redshift in both fields is $z=0.1$}
\tablenotetext{a}{The number of SDSS/MPA galaxies that are \emph{observed} at all $f$ through 24\micron wavelengths (\S\ref{sec:ir_uv}).}
\tablenotetext{b}{The number of observed SDSS/MPA galaxies with fluxes brighter than the noted AB magnitude.}
\tablenotetext{c}{The number of observed SDSS/MPA galaxies that pass the flux, redshift, and size cuts given in \S\ref{sec:limits}.}
\end{deluxetable}

\subsection{Optical Data and Derived Parameters: SDSS}
\label{sec:data_opt}
A description of the SDSS optical reductions is given in \citet{sdss_edr} and \citet{sdss_dr2}.  Throughout this study we use the Petrosian magnitudes supplied by the SDSS Data Release 4 (DR4), as given in the photometric catalogs provided by the MPA/JHU group for SDSS studies\footnote{http://www.mpa-garching.mpg.de/SDSS/}.  Additional measurements of emission-line strengths and spectroscopic index values have been made by the MPA/JHU group using special-purpose code to improve continuum subtraction, and are available from the MPA/JHU Web site.  We have removed duplicate galaxies in these catalogs.

Additional parameters including star-formation rates (SFRs) and stellar masses have also been derived and made publicly available for many of these galaxies by \citet{kauffmann03a} and \citet{brinchmann04}. The stellar mass estimates are obtained by \citet{kauffmann03a} from fits of a suite of stellar populution synthesis models to observed spectral indices (to obtain the dust free mass-to-light ratio) and colors (to estimate the attenuation) . These stellar masses are consistent with stellar masses derived using a simple magnitude and color prescription as in \citet{bell_dejong}, that are in turn consistent with limits from dynamical masses. The gas-phase metallicities of many of the sample galaxies have been determined by \citet{tremonti04} from emission-line fluxes.  These metallicities are only available for galaxies with significant emission lines that are not classified as AGN \citep[see][for the AGN classification criteria]{kauffmann03b}, and are only defined within the SDSS aperture. 

We only consider galaxies for which a mass has been determined by \citet{kauffmann03a} (hereafter SDSS/MPA galaxies). Readers are encouraged to see \citet{kauffmann03a} for a description of the optical selection.  In total we consider 872 SDSS/MPA galaxies in the Lockman Hole that have been observed from the Far-UV through 24 \micron. Of these galaxies, 866 have also been observed at 70 \micron.  In the FLS field we consider an additional 186 galaxies observed in the UV (with sufficient UV effective exposure time, \S\ref{sec:ir_uv}) through 24\micron.

\subsection{IR and UV Data}
\label{sec:ir_uv}
The Lockman Hole field has been observed by \emph{GALEX} in 26 0$\fdg$6 radius circular tiles.  The effective exposure times of these tiles are $\sim 10-15$ks in $f$ and $\sim 10-60$ks in $n$. In the FLS field the effective exposure times are highly variable from tile to tile, and we require exposure times $>5$ks in both $f$ and $n$. The UV photometry is taken from the pipeline-produced catalogs provided by the \emph{GALEX} science team \citep{galex_pipe}.  For each SDSS/MPA galaxy we search the catalogs for objects within 3\arcsec of the SDSS position, choosing the closest object in the rare case of multiple matches.  Since the \emph{GALEX} fields overlap, some objects appear in two separate fields.  We choose the tile with the longer effective exposure time (that takes into account sensitivity variations in the detector).  We also exclude objects $>0.55\deg$ from the \emph{GALEX} field center as the incidence of artifacts is much higher here.  The UV photometry, the principal UV data used in this study, is computed in elliptical Kron apertures using a version of SExtractor\citep{sextractor} modified to operate on low-background images.  

The IR data are from publicly available \emph{Spitzer} imaging. In the Lockman Hole we use the reduced images provided by the \emph{Spitzer} Wide-area Infrared Extragalactic Survey \citep[SWIRE;][]{swire} team\footnote{http://swire.ipac.caltech.edu/swire/astronomers.html for a description of the SWIRE image processing}. In the FLS field we use the post-basic calibrated data (post-BCD) data provided by the \emph{Spitzer} Science Center \footnote{see http://ssc.spitzer.caltech.edu/postbcd/ for a description of post-BCD processing}.  We have checked the consistency of these data sets by comparing the post-BCD Lockman Hole mosaics to the SWIRE-processed data in this field; the photometry is consistent to better than 0.1 mag at the flux levels we consider, except at 70 \micron~(which we do not consider for the FLS field) and at the faintest 24\micron~fluxes where the (random) differences rise to $\sigma \sim 0.5$ mag at $m_{24}=19$ mag. We have performed aperture photometry in the \emph{Spitzer} 3.6 through 8\micron~IRAC images and 24\micron~MIPS images at the location of each of the SDSS/MPA galaxies, using a 7\arcsec radius aperture (12\arcsec at 24 \micron).  The IRAC fluxes are then aperture-corrected to 12.2\arcsec, the radius used for IRAC standard star flux measurements. At 24 \micron~we attempt to calculate total fluxes,  by applying an aperture correction to very large radius ($>35"$) derived from the brightest 24\micron~sources. We have checked that the 7\arcsec radius aperture (12\arcsec at 24\micron) does not induce significant aperture effects by comparing to fluxes in larger radius apertures as a function of $r$ band Petrosian radius; no significant trend is seen for the galaxies of our final sample. At 70\micron~our sources are unresolved, and following \citet{mips70} we apply an aperture correction of 2.07 to fluxes obtained with a 16\arcsec radius aperture with 18-39\arcsec sky annulus.   Systematic errors in IR flux due to calibration uncertainty, aperture corrections, and the resolved nature of many of the sources at wavelengths less than 70\micron~amount to $\sim 30\%$. 

\subsection{$K$-Corrections and Luminosities}
\label{sec:kcorr}
The resulting UV through 3.6\micron~magnitudes are $K$-corrected to $z=0.1$, the median redshift of the sample (e.g. $^{0.1}u$, $^{0.1}g$, etc.), using the method of \citet{blanton_k}. The generalized UV luminosities are then calculated as $L_{uv}=4\pi D_{L}^{2}f_{\lambda}\lambda_{f}$, where $D_L$is the luminosity distance\footnote{Throughout this study we assume a concordance cosmology: $\Omega_{m}=0.3, \Omega_{\Lambda}=0.7, \mbox{H}_{0}=70$ km s$^{-1}$ Mpc$^{-1}$.} and $f_{\lambda}$ is the $K$-corrected flux density per unit wavelength in the $^{0.1}f$ band, with effective wavelength $\lambda_f=1390\AA$.  At longer wavelengths dust emission becomes more important than stellar emission, and we use a different method to ``$K$-correct'' the data:  we choose the best-fitting redshifted \citet{dale01} model IR spectral energy distribution (SED), on the basis of the observed 8 to 24\micron~flux ratio, after correction for stellar emission at 8 and 24\micron~determined from the 3.6\micron~flux following \citet{helou04} and \citet{dale05}.  This best-fit SED is then normalized using the measured 24\micron~flux, and the integrated far-IR (8-1000\micron) dust luminosities ($L_{dust}$) are derived.  Note that the different \citet{dale01} SEDs have $L_{24\micron}/L_{dust}$ ratios that are different by a factor of up to five. We have checked that our results would not change significantly if we use the model SEDs of \citet{devriendt} \citep[see][for a detailed discussion of predicting IR luminosities from \emph{Spitzer} data]{papovich02}. 

\begin{figure*}[t]
\epsscale{0.7}
\plotone{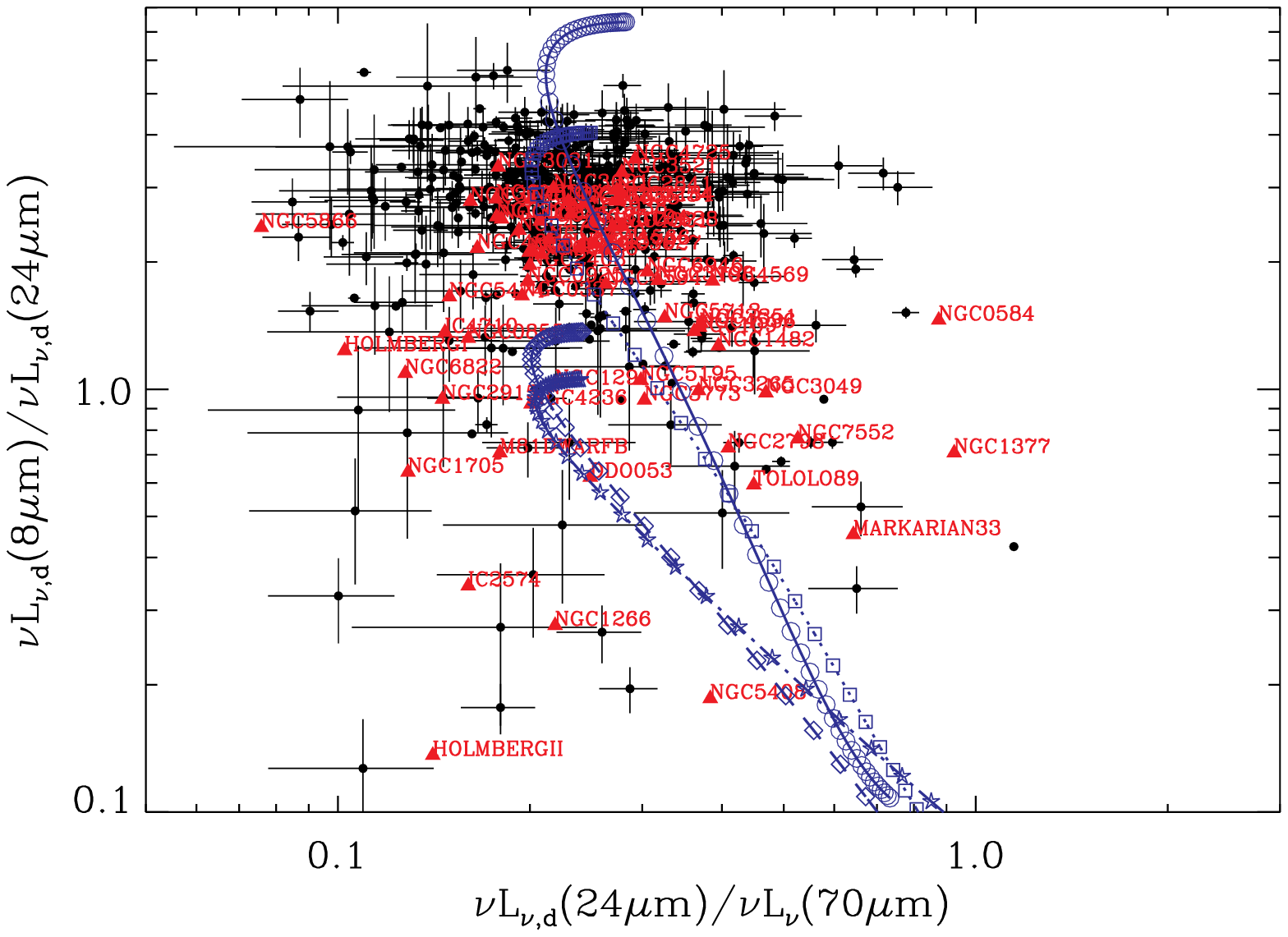}
\figcaption{FIR color-color diagram for the sample with 70\micron~AB magnitude $m_{70\micron} < 15$.  Also shown are integrated values from \protect\citet{dale05} SINGS galaxies ({\it red triangles}) and the models of \protect\citet{dale01} ({\it blue line, diamonds}).  The \protect\citet{dale01} models are shown for redshifts of $z=$ 0.0 ({\it open circles}), 0.1 ({\it squares}), 0.2 ({\it diamonds}), and 0.3 ({\it stars}). The median redshift of the sample is $z=0.1$. Error bars show the photometric measurement errors.
\label{fig:ir_color}}
\end{figure*}

In Figure \ref{fig:ir_color} we show the FIR flux ratios for the sample galaxies, compared to the models of \citet{dale01} and data from the \emph{Spitzer} Infrared Nearby Galaxy Survey (SINGS) sample of galaxies \citep{kennicutt_sings, dale05}.  Note that the 24\micron/70\micron~color does not vary as much as the 8\micron/24\micron color for these models and is not monotonic.  The 24\micron/70\micron~ratio for the sample galaxies, and for the SINGS galaxies, show a larger range of $L_\nu(24\micron)/L_\nu(70\micron)$ than the \citet{dale01} models. Many of the galaxies in our sample are bluer in the 8\micron/24\micron~color than any of the \citet{dale01} models at their redshift, likely due to incomplete treatment in these models of PAH emission, which is assumed to be constant in spectral shape \citep[see][for evidence of changes in the relative strength of PAH features]{smith06}.  In general the scatter observed in this diagram suggests that more complicated and multi-parameter models may be necessary \citep{smith06,draine_li06}. The use of such models with SDSS/MPA galaxies will be explored in a future work \citep{ir_sed}. Some care must be taken in interpreting the total dust luminosities we derive, since these rely on models that do not appear to match existing data to high accuracy.

\subsection{Samples and Limits\label{sec:limits}}
Our primary sample then consists of these galaxies with $m_{24}<19.5$(AB) and $f$ and $n < 25.0$ (AB). It has been found that systematic errors in the $n$ band photometry due to source blending can occur at $n>23$ -- only 45 of our galaxies are at these magnitudes.   The SDSS spectroscopic selection, from which our sample is drawn, also requires $r<17.7$ (AB).  In the Lockman Hole, 80 of the 872 multiwavelength observed SDSS/MPA galaxies fail the UV cut, and 53 galaxies fail the 24\micron~cut, of which 12 also fail the UV cut.  We further require $z>0.02$ (8 galaxies removed) and an $r$ band Petrosian radius (R$_{p}(r)$) of less than 11\arcsec (an additional 23 galaxies removed), to avoid large galaxies which are difficult to accurately photometer.  This leaves a total of 721 SDSS/MPA galaxies out of 872 observed at all wavelengths ($f$ through 24\micron) in the Lockman Hole (82\%).  The galaxies that are missed due to the UV and IR flux cuts are primarily red sequence or early type galaxies (of which $\sim 70$\% \emph{are} detected), although several dwarf galaxies are also missed.  The size cuts primarily remove blue dwarf galaxies -- which tend to be at low-redshift in the SDSS sample -- although a number of dwarfs remain.  In the FLS these same limits yield 118 galaxies detected at all wavelengths out of of 186 observed ($\sim63$\%).  The lower detection rate in the FLS can be explained by the different UV exposure time distribution and the variation of the redshift distribution between the FLS and Lockman Hole fields (which both have median redshift $z=0.1$).

\subsection{Classes of Galaxies}
\label{sec:classes}
We define several special classes of galaxies to be used throughout this study - these follow common definitions in the literature and are non-exclusive.

\begin{center}
\begin{itemize}

\item Dwarfs, with $\log\mbox{M}_*<9$, where M$_*$ in solar units is from \citet{kauffmann03a}

\item Luminous infrared galaxies (LIRGs), with L$_{dust}>10^{11} L_{\odot}$

\item Ultraviolet luminous galaxies (UVLGs), with $\nu L_{\nu}=$ L$_{uv} > 10^{10} L_{\odot}$.  Note that this definition does not include structural parameters (e.g. compactness) which appear necessary in identifying galaxies that may serve as low redshift analogs for Lyman break galaxies (LBG) \citep{heckman05_uvlg}.

\end{itemize}
\end{center}

We also consider the properties of galaxies that have been observed both as part of the SINGS program\citep{kennicutt_sings} and by \emph{GALEX} as part of the Nearby Galaxy Survey \citep[NGS;][]{gil_de_paz}.  The integrated fluxes of these nearby, resolved galaxies are given by \citet{dalex}.  The 65 SINGS galaxies that we consider have publicly available drift-scan spectroscopy as well \citep{moustakas}, from which we have calculated a 4000\AA~break measure, \dfn, for comparison with the SDSS \dfn~values (see \S\ref{sec:dfn}).

\section{The Global Properties of Galaxies in the UV and IR}
In \citet{dissecting} we introduced an observational diagnostic of galaxy properties that relates the SFH and attenuation of a galaxy to its color.  The SFH and attenuation are the primary drivers of galaxy color (although metallicity can play a significant role, especially in older galaxies).  A combination of IR and UV observations provides a measurement of obscured and unobscured SF, and is thus ideal for measuring SFR and attenuation. A detailed exploration of this relation is given in \S\ref{sec:dac}. Here we investigate the location of nearby, resolved galaxies, special classes of galaxies, and the sample of all detected galaxies in the CMD and the relation between color, SFH, and attenuation. These additional diagnostic relations are the CMD and the relation between attenuation and SFR . The locations of the special classes of galaxies (\S\ref{sec:classes}) both places them in the context of the larger, more general galaxy distribution and informs the interpretation of the relations.  The measures of SFH and dust attenuation that we use are the strength of the 4000\AA~ break and the ratio of IR to UV luminosities, and are described here.

\subsection{\dfn}
\label{sec:dfn}
As a measure of SFH we use the definition of the 4000\AA~ break strength given in \citet{balogh99} (\dfn),  since it is measurable at high signal to noise for the galaxies that we consider (i.e both red and blue sequence galaxies), and is nearly insensitive to dust reddening \citep{macarthur05}. The 4000\AA~ break is similar to a measure of the ratio of the SFR averaged over $\sim 10^{8.5}$ yr to the SFR averaged over $>10^{9}$ yr but is subject to strong variations with metallicity, especially at larger break strengths.  In \S\ref{sec:dac} we explore the behavior of \dfn~for some simple galaxy star formation histories using stellar population synthesis models at two different metallicities. While there are some variations with metallicity, in general \dfn~increases for smaller ratios of current to past averaged SFR (i.e ``older'' galaxies). One complication in the use of \dfn~is that it is only measured within the spectroscopic aperture, and so galaxies with moderate bulge/disk ratios may have \dfn~overestimated -- this is clearly the case for a small number of galaxies in this sample (see also \citet{kauffmann06}).

\subsection{IRX}
\label{sec:irx}

A common and robust attenuation indicator is the ratio of IR dust emission to UV stellar emission (the so-called infrared excess, IRX$=\log (\mbox{L}_{dust}/\mbox{L}_{uv})$).  It is independent of and a qualitatively different kind of measure than various optical measures based on a color excess (e.g. the Balmer decrement). This indicator is predicated on the idea that the UV emission measures the amount of transmitted flux from young stars while the IR emission (here $\lambda\lambda~8-1000\micron$) measures the amount of UV flux from young stars absorbed by dust and re-radiated in the thermal IR \citep{buat92, calzetti94, kennicutt98, gordon00}. Under these simplistic assumptions the attenuation in the UV (in magnitudes) can be written $A_{uv}=2.5\log (10^{IRX}+1)$. Light will be absorbed at wavelengths other than the UV, but the combination of dust attenuation curves that rise steeply towards the blue with blue intrinsic spectra has justified these assumptions when considering intensely star-forming galaxies.  For many of the galaxies we consider this approximation may not be valid -- specifically the fraction of the IR luminosity due to the heating of dust \emph{by UV light} may vary between galaxies --  but following \citet{MHC} we define 

\begin{equation}
\label{eqn:airx}
\widehat{A}_{IRX}=2.5\log (\eta 10^{\mbox{IRX}}+1) 
\end{equation}
with $\eta=1/1.68$ for use when comparing the IRX we determine to other measures of attenuation in magnitudes. The parameter $\eta$ gives the fraction of the IR luminosity that comes from dust-absorbed UV light produced by young stars.  Note that other conversions between IRX and attenuation are available \citep[e.g.,][]{bell02b, buat05} but are generally consistent with this formulation, which has the advantage of being physically motivated and simply understood.  The value of $\eta$ that we choose is similar to that chosen by \citet{buat07} and is consistent with \citet{MHC}, \citet{bell03_radio}, \citet{hirashita03}, and \citet{iglesias_paramo06}.  In \citet{paper2} we attempt to estimate $\eta$ for each galaxy.

In detail several modifications to the naive interpretation of IRX as a measure of attenuation are possible.  First, there is the aforementioned effect of some fraction of the light being absorbed at wavelengths other than $\lambda\sim 1400$\AA, which will vary depending on the SFH of the galaxy. Second, the possibility of different stellar populations within the galaxy being attenuated by different amounts of dust must be considered \citep{charlot00}.  A third and related complication is the effect of the relative geometry of the stars and the dust, on both large and small scales \citep{WG00,buat96}, although \citet{gordon00} find that for young stellar populations IRX is a good measure of the UV attenuation for a range of dust geometries and extinction laws, but they do not consider the effects of inclination in realistic disks on this measure \citep[e.g.,][]{pierini04}. Finally, older, blue horizontal branch stars may contribute to the UV flux, and AGN may contribute to both the UV and IR emission.  However, despite these complications it appears that, at least for galaxies with \dfn$<1.7$, IRX is an accurate dust estimator that increases monotonically with $A_{fuv}$ for a given SFH \citep{paper2}. 

It is important to clarify the meaning of IRX as used throughout this paper. For simplicity, and to conform with the convention in the literature, we will use IRX defined by the total IR and UV luminosities  (IRX$=log(\mbox{L}_{dust}/\mbox{L}_{UV})$, where L$_{dust}$ and L$_{UV}$ are defined in \S\ref{sec:kcorr}). The issues mentioned above regarding the relation of IRX to the true FUV attenuation are considered separately in \citet{paper2}.

\subsection{Color-Magnitude Diagram}
\label{sec:cmd}

The CMD for galaxies is a fundamental observational tool in the study of galaxy evolution. Neglecting the effect of dust (and to a lesser extent metallicity), the color coarsely measures the SFH of a galaxy, while the absolute magnitude in one of the redder bands measures the total stellar mass (again subject to the effects of dust attenuation, and and to a lesser extent the SFH).  The CMD thus relates the current star-formation to the past-integrated star formation of galaxies.  \citet{baldry} and  \citet{blanton03_prop} definitively identified the bimodality of galaxies in this diagram at low redshift using the SDSS, with many galaxies having red or blue colors and few galaxies at intermediate color. \citet{kauffmann03a} and \citet{brinchmann04} have constructed the SFH and stellar mass analog of the CMD for $\sim 500,000$ SDSS galaxies. Studies of the evolution of the distribution of galaxies in the CMD \citep{bell04_red, faber05} have identified an increase in the mass density of the red sequence from $z\sim0.7$ to $z\sim 0.1$.

\begin{figure}[t]
\epsscale{1.0}
\plotone{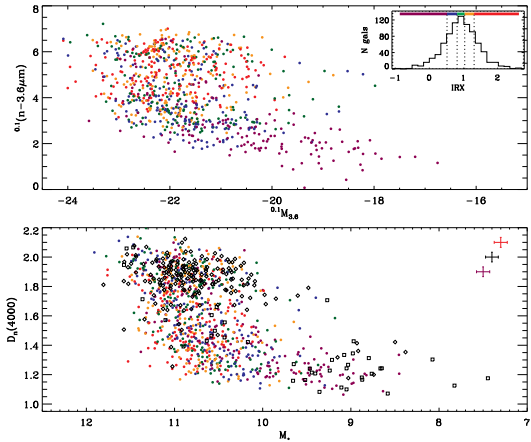}
\figcaption{The color magnitude diagram for the sample detected at all wavelengths.  {\it Top}: The $^{0.1}(n-3.6\micron)$ color vs. $M_{3.6}$.  The color of each symbol encodes its IRX quintile, from lowest (purple) to highest (red).  The median photometric errors are smaller than the symbol size.  The inset shows the distribution of IRX and the bins used to color-code the symbols.  {\it Bottom}: Same as top, but with \dfn~as the $y$-axis and M$_*$ from \protect\citet{kauffmann03a} as the $x$-axis. Black diamonds show galaxies undetected in the UV or at 24\micron, while squares mark galaxies with $z<0.02$ or Petrosian radius greater than 11\arcsec in the $r$ band.  Error bars show the median error for the lowest quintile ({\it purple}) and highest quintile ({\it red}) of IRX, as well as the median error for the entire sample ({\it black}).
\label{fig:cmd}}
\end{figure}

In Figure \ref{fig:cmd} we show the CMD of the sample galaxies.  In the top panel we use the $n-3.6$\micron~color - this is similar to the traditional CMDs using $u-r$ or $g-r$ \citep{baldry, blanton03_prop}, except that the use of the $n$ magnitude causes greater separation of the red and blue sequences \citep{wyder07}.  The color of each symbol encodes its IRX quintile: red symbols are in the highest IRX quintile and tend to be located near the massive, high \dfn~edge of the blue sequence or on the red sequence, while purple symbols are in the lowest IRX quintile and are found primarily in the low mass tail of the blue sequence (i.e., dwarfs). In the botom panel we use \dfn~as the measure of SFH.  This measure is more directly related to SFH than the broadband color - as we will show below the color is additionally affected by dust attenuation.  The stellar mass estimates are from \citet{kauffmann03a} (\S\ref{sec:data_opt}).  In this diagram we also show the location of galaxies undetected in the UV or at 24\micron~-- these are approximately $\sim 25$\% of the SDSS/MPA galaxies with \dfn$>1.7$, as well as some dwarfs.  Galaxies that do not pass the size cuts ($z>0.02$, R$_p(r)<11$'') are also identified.

We clearly see the red and blue sequences, and their separation, in both diagrams. The blue sequence is tilted such that it becomes redder with increasing stellar mass. The degree of tilt, the scatter around the main sequences, and the evolution of both can be used to constrain the SFH of galaxies \citep{ss92, faber05, noeske07b, schiminovich07}. There is also an increase in the average IRX with mass.  Many of the galaxies in the `green valley' between the red and blue sequence appear to have large IRX. The accurate correction of observed colors for the presence of dust is thus a crucial step in determining the true tilt of the blue sequence for comparison to theories of galaxy formation.  Attenuation is mass (and SFR) dependent, and so inaccurate attenuation corrections can lead to incorrectly determined distributions of SFR with mass.

\subsection{Dust-SFH-Color}
\label{sec:dac_classes}
\begin{figure*}[ht]
\plotone{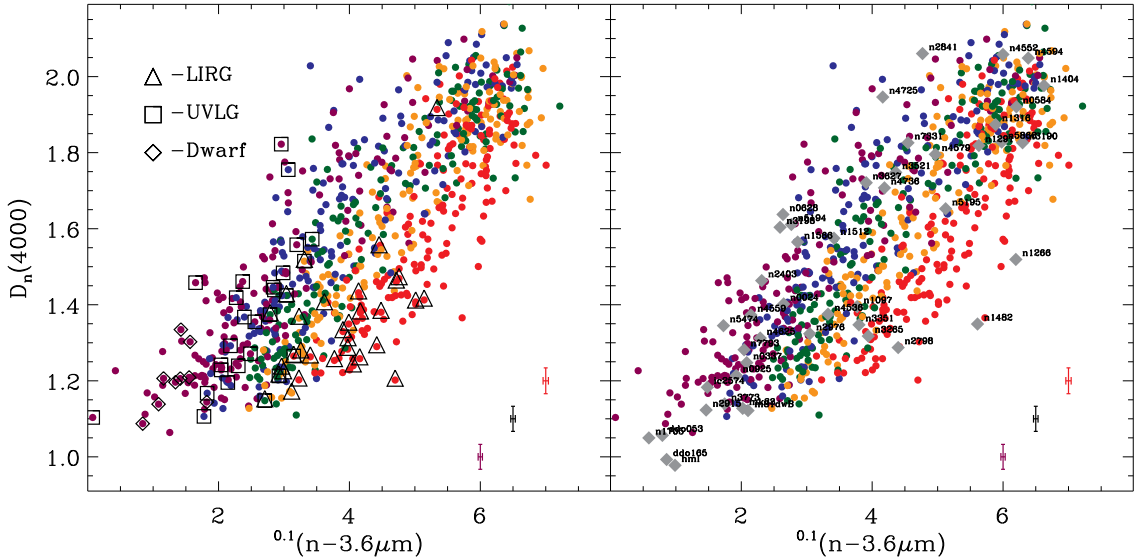}
\figcaption{ {\it Left}:  Relation between IRX, \dfn, and color with location of LIRGs, UVLGs, and dwarf galaxies marked (see \S\ref{sec:classes} for definition of these classes). The color of each symbol indicates the IRX quintile, from lowest ({\it purple}) to highest ({\it red}) as in Figure \ref{fig:cmd}. Error bars in the lower right show the median error for the lowest quintile ({\it purple}) and highest quintile ({\it red}) of IRX, as well as the median error for the entire sample ({\it black}). {\it Right}: Same as left, but with the location of galaxies from the SINGS and NGS surveys marked by grey diamonds.
\label{fig:classes}}
\end{figure*}

In Figure \ref{fig:classes} we show the relation between IRX, \dfn, and $n-3.6$\micron~color for the sample of galaxies presented in this work.  The color of each symbol indicates the IRX quintile.  This diagram was introduced in \citet{dissecting}. There is a clear relation between \dfn~and color for the lowest quintiles of IRX (i.e., purple or blue symbols), though the relation becomes somewhat more scattered at high \dfn.  For larger values of IRX (i.e. orange or red points) the relationship between color and \dfn~persists, but is shifted to redder colors because of the reddening effect of the dust.  Figure \ref{fig:classes} shows how the color is additionally affected by the attenuation as well as SFH -- in this diagram the reddening vector is nearly horizontal -- and how the conversion from color to \dfn~in the CMD may be made. (i.e., to go from the top to the bottom panel in Figure \ref{fig:cmd}).  

The locations of LIRGs, UVLGs, and dwarfs are shown in Figure \ref{fig:classes}$a$ while the locations of nearby galaxies drawn from the SINGS and \emph{GALEX} NGS are shown in Figure \ref{fig:classes}$b$.   The LIRGs clearly occupy the the reddest region for a given \dfn, and typically follow the sequence of galaxies with large IRX.  The UVLGs are generally blue for their \dfn, and have relatively low attenuation although they do not follow the sequence of galaxies with the \emph{lowest} IRX. The nearby galaxies cover the parameter space probed by the current sample reasonably well. 

\begin{figure*}[t!]
\plotone{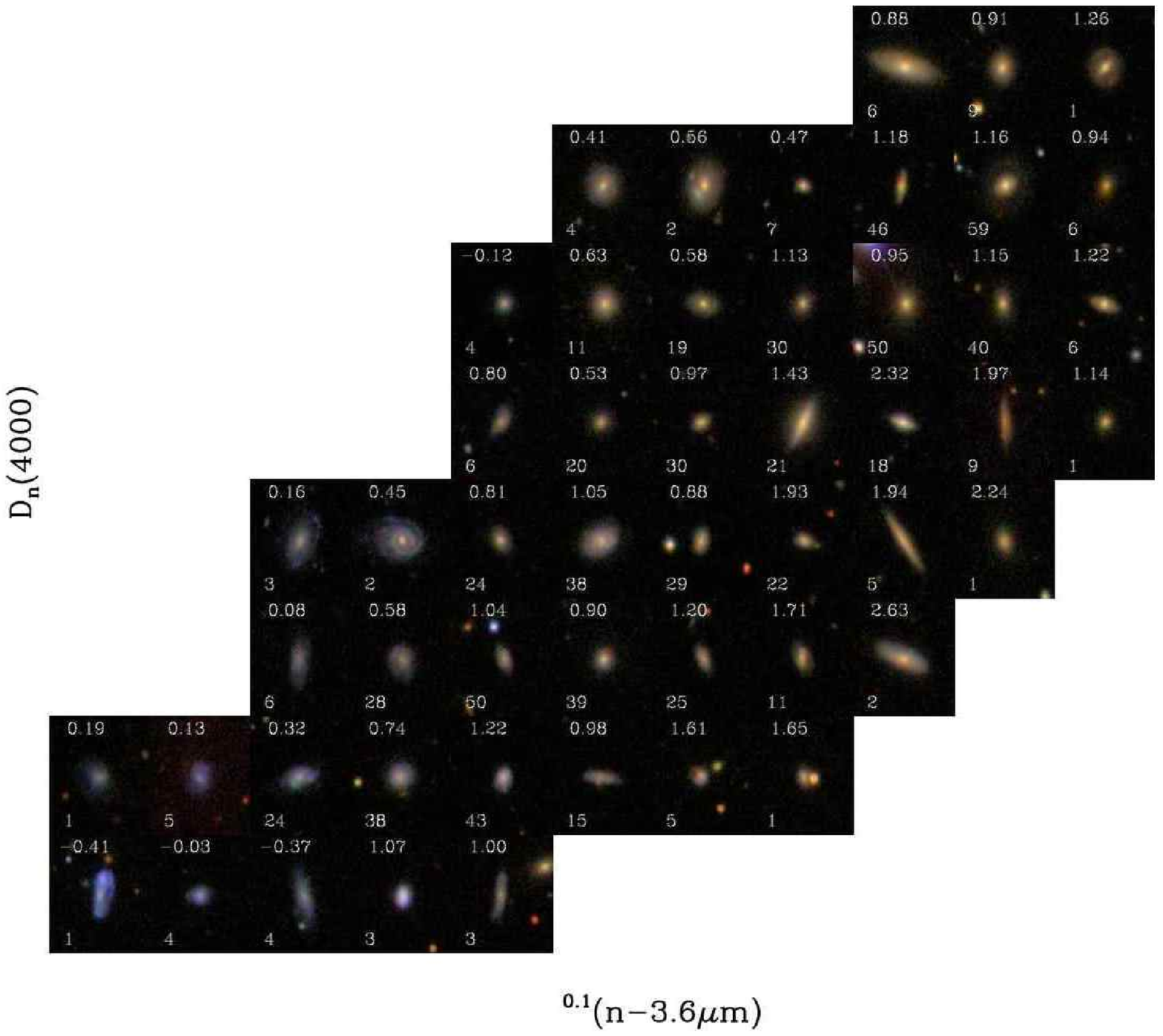}
\figcaption{ SDSS $gri$ color composite images of galaxies, organized in bins of \dfn~and $^{0.1}(n-3.6\micron)$ color.  For each bin of \dfn~or color (one eighth and one eleventh of the range, respectively) a galaxy is chosen at random for display.  The top number in each image indicates the median IRX value for that bin, while the bottom number gives the number of galaxies in the bin.
\label{fig:morph}}
\end{figure*}

In Figure \ref{fig:morph} we show SDSS $gri$ composite images of galaxies in the sample, selected randomly from bins of \dfn~and $^{0.1}(n-3.6\micron)$ color.  At high \dfn~and relatively blue color we can see examples of galaxies with red bulges and blue, star forming discs, highlighting the effect of aperture on the \dfn~measurement \citep{kauffmann06}.  At moderate \dfn~and very red color there are many highly inclined, apparently red, discs, showing the effect of attenuation on the color while \dfn~is relatively unaffected.

\section{Implications of the Relation Between Attenuation, SFH, and Color}
\label{sec:dac}

Having identified the relation between dust attenuation, SFH, and color as a useful diagnostic of galaxy properties, here we seek to explore the nature of the relation in more detail by considering different projections and a wider range of colors sampling the entire UV-optical SED. We also compare the relation to models that predict, separately, the effects of dust attenuation and SFH on the color. In this way we can determine which UV-optical colors, in combination with \dfn~and IRX, best constrain the models of attenuation and SFH. Finally, we parameterize the empirical relationship for these additional colors and examine the accuracy of the parameterization, and the degree to which deviations are correlated with several additional physical properties of the galaxies.

\subsection{Extension to Additional Colors}
\begin{figure*}[t]
\epsscale{1.0}
\plotone{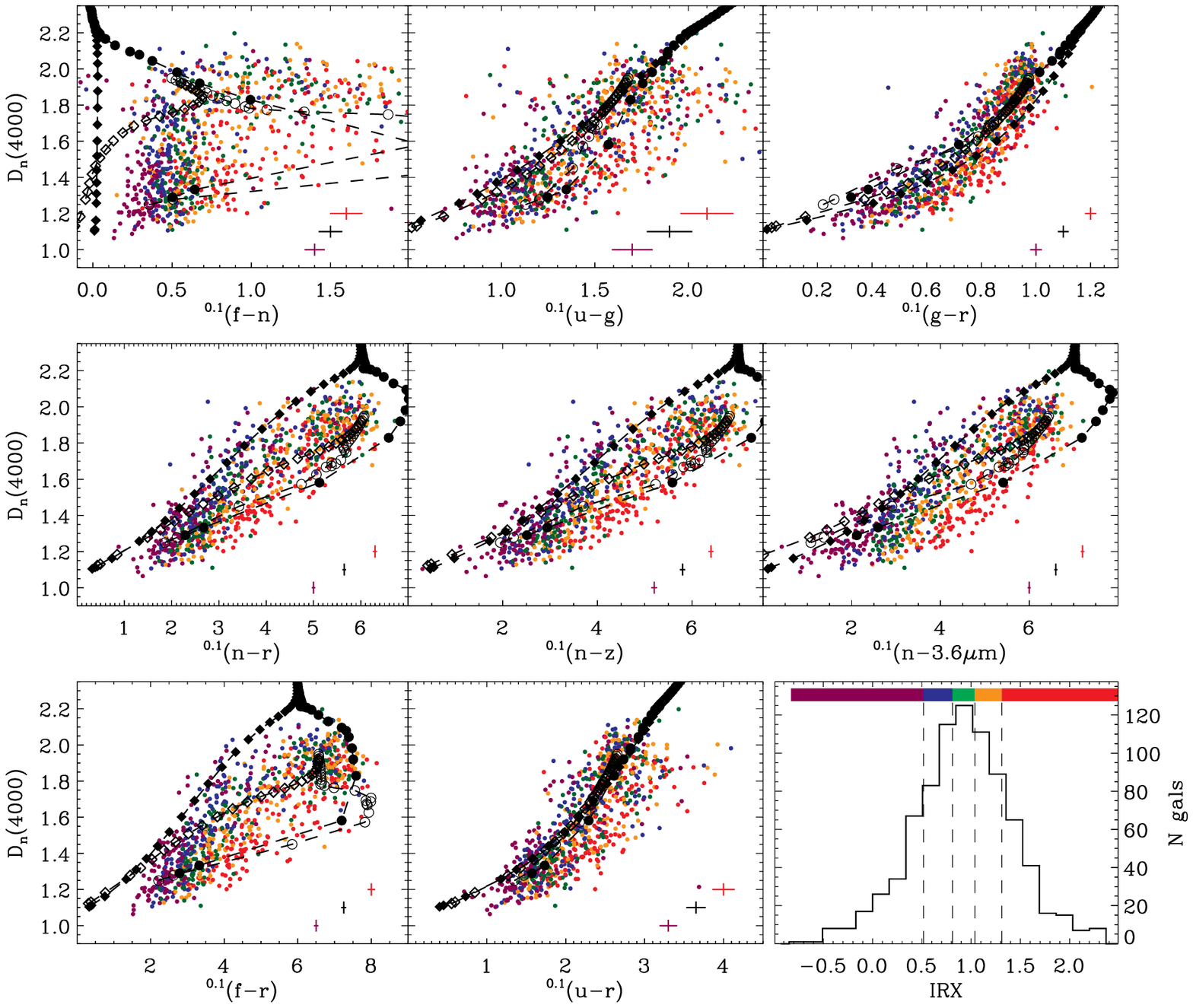}
\figcaption{Relation between \dfn~and color for different bins of IRX, for the colors $f-n,u-g,g-r,,n-r,n-3.6,f-r,u-r$.  The bins of IRX used to color code the symbols are shown at lower right.  Error bars in the lower right of each panel show the median error for the lowest quintile ({\it purple}) and highest quintile ({\it red}) of IRX, as well as the median error for the entire sample ({\it black}). Overplotted in black are four \protect\citet{BC03} population synthesis models with Z/Z$_{\sun}=0.4$ ({\it open symbols}), $2.5$ ({\it filled symbols}) and exponentially declining SFR with time constants $\tau_{SF}=0.1$ Gyr ({\it circles}), $1$ Gyr ({\it diamonds}). The symbols are placed at intervals of 500 Myr, from 0.5 to 13 Gyr.
\label{fig:dfn_color}}
\end{figure*}

In Figure \ref{fig:dfn_color} we show \dfn~versus color for a number of different broadband colors. The top panels show colors with a short wavelength separation, while the middle panels show colors with a longer wavelength separtion.  As in Figure \ref{fig:classes} the symbols are color-coded by their IRX quintile. We see a clear separation of the effects of dust and SFH on the colors with widely separated wavelengths.  This behavior is discernible in the colors with a shorter wavelength separation as well (e.g. $g-r$, $u-g$), but the separation is not as clear. For each panel of Figure \ref{fig:dfn_color} the effect of dust attenuation is to move galaxies nearly horizontally along the color axis, since the effect of attenuation on \dfn~is minimal (but see \S\ref{sec:bycolor}). In effect, the 'reddening' vector is nearly horizontal. Thus a relation between \dfn~and color that appears for galaxies with low IRX ({\it purple symbols}) will simply be translated along the color axis for galaxies with large IRX ({\it red symbols}) if the relation between \dfn~and unattenuated color is nearly constant among galaxies.  The size of the translation is constrained by IRX.  The clear separation of \dfn~and dust effects also implies a nearly invariant relation between \dfn~and unattenuated color, at least for regions of the diagram with low scatter (\dfn$<1.7$).

There are some changes in the form of the dust-SFH color relation depending on the color used.  In particular, at red $g-r$ color ($g-r \sim 0.9$) there is saturation in the sense that $g-r$ color does not change significantly even as \dfn~increases or for different values of IRX. This is in contrast to the $n-r$ and other long-wavelength baseline colors where there is a relatively large variation in color for different values of \dfn~and IRX.  At low \dfn~there is a difference in behavior of the $n-r$ and $n-3.6$\micron~colors.  The $n-3.6$\micron~color shows a larger range at low \dfn~than the $n-r$ color -- this is because for a given \dfn~and large dust attenuations the $r$ band flux is affected more than the 3.6\micron~flux, and so the $n-r$ color does not increase as rapidly as the $n-3.6$\micron~color for increasing attenuations.

The trend of \dfn~with $f-n$ color for a given narrow range of IRX is related to the effect of SFH on the UV spectral slope $\beta$ as described by \citet{kong04}.  Here we see that there is very little, if any, correlation of $f-n$ with \dfn~for \dfn$<1.8$ -- bands of nearly constant IRX appear vertical in the \dfn-color plane, although there is a clear shift to redder colors for larger values of IRX. The IRX-$\beta$ relation is considered in more detail in \citet{paper3}. 

At \dfn$>1.8$ the relation between \dfn~and color is different -- there is a ridge of galaxies for which $f-n$ \emph{decreases} for increasing \dfn.  IRX appears to be significantly less correlated with $f-n$ for these galaxies than for those with \dfn$<1.8$.  As we see below (\S\ref{sec:bc03}) this ridge is close to the path followed by galaxies with a burst-like SFH and a small ratio of current to past averaged SFR, and traces the increasing importance of \emph{old} blue stars in galaxies with very small ratios of young stars to old stars.

\subsubsection{SFH models}
\label{sec:bc03}

To aid in the interpretation of the relation between \dfn~and SFH -- and between SFH and color -- we have constructed several stellar population synthesis (SPS) models of galaxy spectra using the \citet{BC03} code.  These models use the \citet{chabrier} IMF with the Padova 1994 tracks, and have smooth, exponentially declining SFR$\sim e^{-t/\tau_{SF}}$ with no gas recycling.  Colors (in the blueshifted filters) and \dfn~are calculated directly from the resulting spectra. No dust attenuation has been applied. We show four of these models in Figure \ref{fig:dfn_color}, having metallicity $Z/Z_{\sun}=0.4$ or $2.5$ and $\tau_{SF}=0.1$ or 1 Gyr.  None of the models are star-bursting in the sense that the ratio of the current to past-averaged SFR is always less than 1. A model with $\tau_{SF}=10$Gyr follows tracks similar to the $\tau_{SF}=1$Gyr models but never reaches \dfn~$>1.4$. In this diagram the models with $\tau_{SF}=0.1$ are similar to a simple stellar population (SSP) model. Models with different $\tau_{SF}$ generally occupy the same region of \dfn-color space at late times ($t>12$ Gyr), while models with different $Z$ have significantly different \dfn~at late times.  It is important to keep in mind that the $K$-corrections we have applied are derived from spectra based on the \citet{BC03} models, so some care must be taken in drawing conclusions based on small effects in color. 

The zero-dust SPS models nevertheless yield several interesting insights.  First, the models with different $\tau_{SF}$ trace very different paths in the \dfn~versus $f-n$ diagram.  For the $\tau_{SF}=1$ model the only change in $f-n$ occurs at late times for $Z/Z_{\sun}=0.4$, even as \dfn~steadily increases.  This is consistent with the majority of galaxies at \dfn$<1.7$ that, for a given small range in IRX, show little variation in $f-n$ color even as \dfn~varies.  The $\tau_{SF}=0.1$ model exhibits strong variations in $f-n$ color as a function of time.  The models start at $^{0.1}(f-n)\sim 0.5$ and become very red reaching a maximum of $^{0.1}(f-n)\sim 2$ or $\sim 2.5$ for the high and low-metallicity models, respectively, at $t\sim1.5-2$ Gyr. After this time the models become bluer as the emission from evolved blue stars becomes dominant, while \dfn~continues to increase.  Although several SDSS/MPA galaxies fall along the $\tau_{SF}=0.1$ track at low \dfn~and redder colors, these galaxies also have significant IRX, such that they are incompatible with the \emph{zero-dust} $\tau_{SF}=0.1$ model.  The attenuation corrected $^{0.1}(f-n)$ color of these galaxies would place them closer to the zero-dust $\tau_{SF}=1$ model.  In contrast, many of the galaxies with \dfn$>1.8$ lie along a ridge that appears similar to the track of the fast declining ($\tau_{SF}=0.1$) model, and these galaxies appear to have little relation between their color and IRX.  

For the long wavelength separation colors ($^{0.1}(n-r)$, $^{0.1}(n-z)$, and $^{0.1}(n-3.6$\micron)) the zero-dust $\tau_{SF}=0.1$ and $\tau_{SF}=1$ models follow much more similar tracks in \dfn-color space.  The galaxies in the lower IRX bins ({\it purple and blue symbols}) follow a similar track, although there is more scatter for \dfn$>1.7$.  The galaxies with the largest IRX ({\it orange and red symbols}) are redder than the unattenuated models.  Galaxies with significantly higher \dfn~than any of the models for a given color are likely those for which the aperture effect causes an overestimate of global \dfn, and have moderate bulge-to-disk ratios \citep[e.g.,][]{kauffmann06}.  The increased scatter at \dfn$>1.7$ may be due in part to increased scatter in the relation between IRX and attenuation for such galaxies (\S\ref{sec:irx}).  In addition, and importantly, it is at \dfn$>1.7$ where the models begin to diverge, and our simple relation between \dfn, attenuation, and color may be broken because of scatter in the relation between \dfn~and color. This scatter is in itself interesting, since it arises from different SFHs. It is only with robust, independent estimates of the attenuation that these small differences in the relation between color and \dfn, corresponding to different forms of the SFH, can be determined from observations.  The variation between the models of the relation between \dfn~and unattenuated color for these longer wavelength separation colors suggests that these colors (as opposed to the optical colors, see below) give some clue to the details of the SFH.  This makes the plots involving these colors, and the $f-n$ color, the most useful for investigating the details of the SFH.

For the colors $^{0.1}(u-g)$, $^{0.1}(g-r)$, and $^{0.1}(u-r)$ the behavior of the models is different.  These colors do not completely straddle the redshifted 4000\AA~ break.  For a given \dfn~the different unattenuated models are nearly identical in these optical colors, even at \dfn$>1.7$.  Because the expected, unattenuated relation between \dfn~and color appears nearly independent of the details of the SFH for these colors, deviations from this relation are more reliably interpreted as due to dust attenuation.  Thus, dust attenuation would appear to be more easily inferred using the relation between \dfn~and these colors than for the longer-wavelength baseline colors that sample the UV (e.g, $^{0.1}(n-z)$).   However, these optical colors also have larger errors relative to the length of the reddening vector than the long-baseline colors, which makes the inference of attenuation more difficult.

\subsection{Binned by IRX}
\label{sec:byirx}
\begin{figure*}[t]
\epsscale{1.0}
\plotone{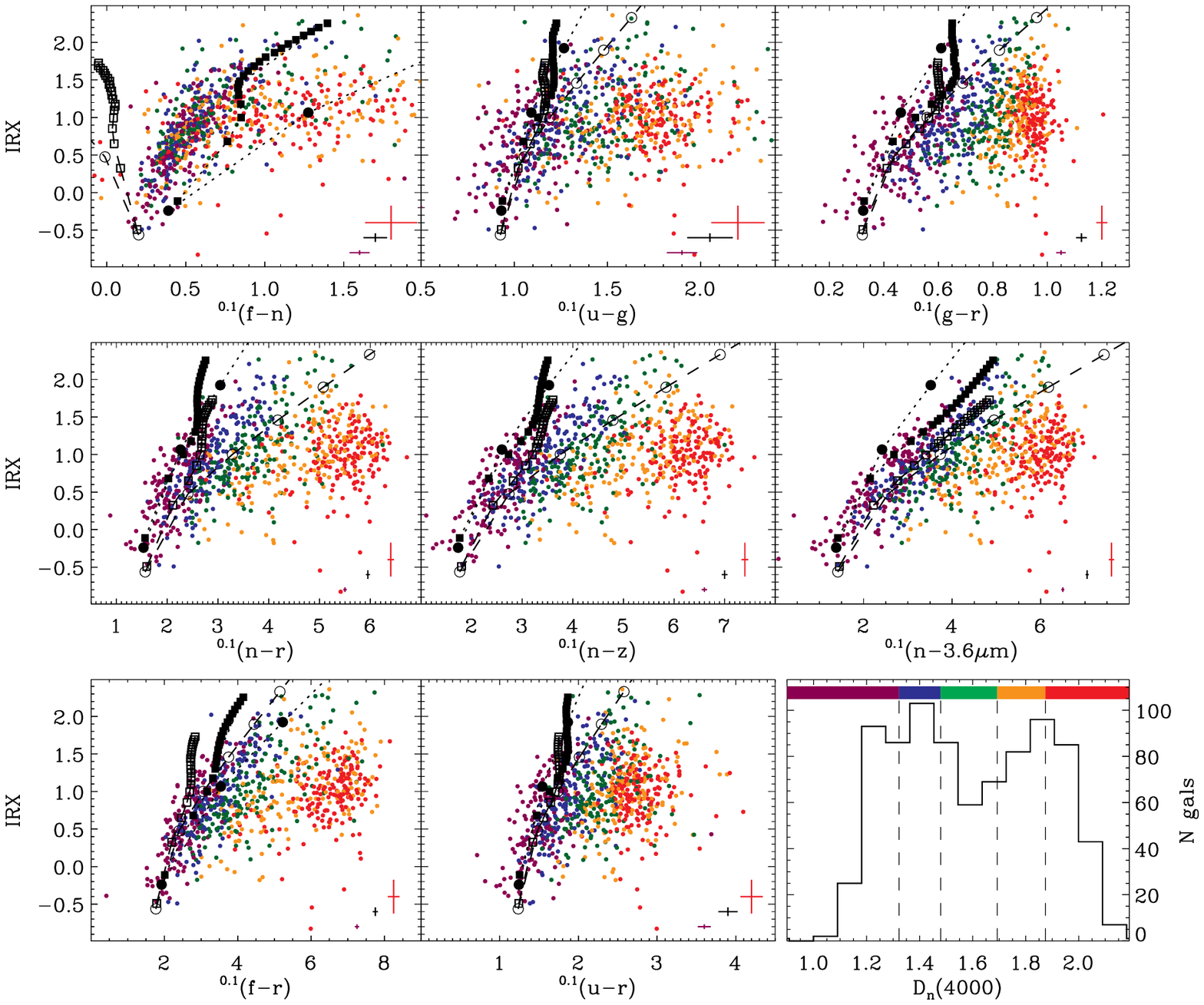}
\figcaption{Relation between IRX and color for different bins of \dfn, for the colors $f-n$, $u-g$, $g-r$,  $n-r$, $n-z$, $n-3.6$\micron, $f-r$, and $u-r$.   The bins of \dfn~used to color code the symbols are shown at lower right. Error bars in the lower right of each panel show the median error for the lowest quintile ({\it purple}) and highest quintile ({\it red}) of \dfn, as well as the median error for the entire sample ({\it black}). The black symbols and dashed lines show the effect of dust attenuation from \citet{WG00}, with a SHELL global geometry and for combinations of a MW ({\it open symbols}) or SMC ({\it filled symbols}) extinction law and a clumpy ({\it squares}) or homogeneous ({\it circles}) local dust distribution.  Symbols run from $\tau_V=0.1$ to $10$ in steps of 0.5.  See text for details. 
\label{fig:irx_color}}
\end{figure*}

In Figure \ref{fig:irx_color} we show the relation between IRX and color for a number of different broadband colors. This is a different projection of the relation shown in Figures \ref{fig:classes} and \ref{fig:dfn_color}, and here the symbols are color-coded by \dfn~quintiles.

For the $f-n$ color the relation between IRX and color is analogous to the IRX-$\beta$ relation \citep{MHC,kong04}, where $\beta$ is the power-law slope of the UV spectrum. The correlation between IRX and $f-n$ has low scatter, and there is very little or no dependence on \dfn~for \dfn$< 1.7$. \citet{kong04} have suggested that the additional effect of SFH on UV color may account for the scatter in the IRX-$\beta$ relation. This would make the UV slope less reliable as a measure of attenuation in, e.g. high redshift galaxies, without knowledge of the SFH. \citet{gordon00} argue that the effects of geometry may make the UV slope a poor measure of attenuation. We find that, for a variety of galaxies with \dfn$<1.7$ observed in the local $z < 0.3$ universe, the $f - n$ color is well correlated with IRX, although the slope is steep and therefore poorly constraining in the presence of significant errors in $f-n$ or $\beta$.  However, for \dfn$> 1.7$ there is a large scatter in the relation, suggesting that the $f -n$ color for these galaxies is not primarily driven by the attenuation (\S\ref{sec:bc03}) or that IRX is not a good measure of the attenuation for these galaxies \citep{paper2}. It is likely that both are true. The IRX-$\beta$ relation for this sample of galaxies is treated in more detail in \citet{paper3}.

The $g - r$ and $u - g$ colors and near equivalents (such as $U -V$) are often used as indicators of red and blue sequence membership in the CMD \citep{bell04_red,faber05}. We see here that they are somewhat affected by dust reddening \citep{bell_gemsdust}, and that care must be taken in defining the red and blue sequence, especially if there is evolution in the distribution of attenuation of galaxies as a function of redshift.

For the long wavelength separation colors (e.g. $n - r$ or $n - 3.6$\micron) it is easy to see the effect of SFH on the color that was predicted by \citet{kong04} for the UV color. For a given low \dfn~(i.e. much recent SF, \emph{purple symbols}) the relation between IRX and color is clear. This relation is closely related to the effective attenuation curve \citep[e.g.,][]{MHC}. As \dfn~increases (i.e. for different colored symbols in Figure \ref{fig:irx_color}) the relation moves to redder color, but the slope and scatter do not change a great deal until the largest \dfn. This shift to redder color at a given IRX is due to the redder intrinsic spectrum of a galaxy with larger \dfn (lower ratio of recent SFR to past averaged SFR). However, at the highest \dfn~the relation between IRX and color becomes more scattered. 

\subsubsection{Dust Models}
\label{sec:wg00}

To more explicitly show the relation of Figure \ref{fig:irx_color} to dust attenuation laws we have overplotted the relation between reddening and FUV attenuation given by the models of \citet{WG00}. These models include several different local and global dust geometries, as well as both Milky Way (MW) and Small Magellanic Cloud (SMC) extinction laws.  The several global dust geometries consist of equally mixed dust and stars in a sphere (DUSTY), a shell of dust surrounding a sphere of stars (SHELL), and a cloud of dust at the center of the more extended stellar distribution (CLOUDY, where the radius of the dust is 0.69 times the radius of the stellar sphere). The local distribution is described as either (h)omogenous or (c)lumpy. To place the model tracks in Figure \ref{fig:irx_color} we convert the attenuation at $\sim$1400\AA~to IRX using equation (1). \citet{gordon00} show that this is a good approximation for all geometries with young stellar populations, although as mentioned in \S\ref{sec:irx} for the older galaxies in the sample this approximation may not be valid. We calculate the color excess by taking the difference of the attenuations at the effective wavelength of the filters. The zero point in color is determined by averaging the color of the five bluest galaxies in the lowest \dfn~quintile. However, for ``older'' galaxies that will have a redder intrinsic, unattenuated color the model curves can be translated along the color axis to match the zero dust color for that galaxy -- in other words, the horizontal zeropoint of these tracks is somewhat arbitrary, depending on the unattenuated spectrum of the galaxy.

For clarity we only show the SHELL geometry in Figure \ref{fig:irx_color}. The CLOUDY model significantly underpredicts both IRX and the color excess for $\tau_{V}<10$. While a few galaxies may be described by this model, the majority do not appear to be in this configuration, for the parameters used by \citet{WG00} ($r_{dust}/r_{stars} = 0.69$). The same is true of the DUSTY model with a clumpy local geometry.  The MW dust extinction law is in significant conflict with the observed IRX versus $f - n$ color relation.  Because of the 2175\AA~bump in the MW extinction law \citep{dust_bump, draine_li06} the $f-n$ color becomes \emph{bluer} with increasing $\tau_{V}$, a behaviour not observed in the data. As can be seen in the top left panel of Figure \ref{fig:irx_color}, and as has been noted by other authors, the relation between UV slope and IRX is a very strong constraint on models of dust attenuation, particularly the extinction law. Note that some models that fit the data well for a given color do not fit the other colors. We find that the SMC extinction law with the SHELL geometry with clumpy or homogenous local geometries, or the DUSTY model with homogenous local geometry provide the best matches to the trends of IRX with color excess in the data, although the DUSTY model gives significantly lower IRX and color excess for a given $\tau_{V}$ than the SHELL model.

\subsection{ Binned by Color}
\label{sec:bycolor}

\begin{figure}[h!]
\epsscale{1.0}
\plotone{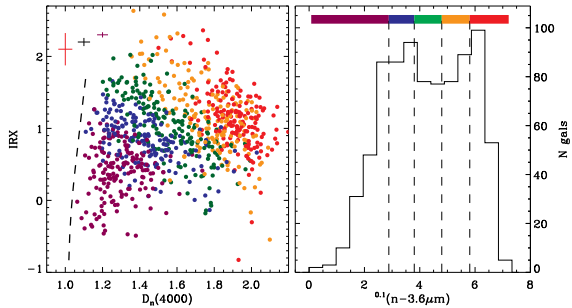}
\figcaption{{\it Left:} Relation between IRX and \dfn~for different bins of $^{0.1}(n-3.6\micron)$ color. The dashed line shows the effect of attenuation on \dfn~for one of the models of \citet{macarthur05}.  Error bars in the top left corner show the median error for the lowest quintile of $^{0.1}(n-3.6\micron)$ ({\it purple}) and highest quintile ({\it red}), as well as the median error for the entire sample ({\it black}). {\it Right}: Histogram of $^{0.1}(n-3.6\micron)$ color, showing the quintiles used to color code the symbols at left.  
\label{fig:dfn_irx}}
\end{figure}

In Figure \ref{fig:dfn_irx} we show the third projection of the dust-SFH-color relation, IRX versus \dfn~for different ranges of $n - 3.6$\micron~color. This effectively shows the region of IRX and \dfn~parameter space allowed for a given color. The extremely small amount of overlap of different color bins suggests that galaxies follow nearly parallel `isocolor' lines in this diagram. Note that at high \dfn~such isocolor lines do overlap, suggesting that an additional parameter may be necessary to explain the $n - 3.6$\micron~color of these galaxies.  This diagram also shows the correlation (or lack thereof) between IRX and SFH in galaxies, and is comparable to similar diagrams constructed for resolved galaxies in the local universe \citep{dalex}.  However, one must again be cautious due to the effect of SFH on the interpretation of IRX as a measure of attenuation.

The envelope of lowest \dfn~galaxies moves to slightly higher \dfn~at higher IRX, so that there are no galaxies with significant dust attenuation and extremely low \dfn. It is possible that galaxies undergoing recent star-formation in the local universe rarely have much dust. However, it is also quantitatively consistent with the increase in \dfn~expected due to dust reddening, an indication that \dfn~is not a perfectly `clean' indicator of SFH \citep{macarthur05}. Using an approximation for effective optical depth,  $\tau_V = [(5500/1390)^{-0.7}/\log e]\cdot$\airx~ where we assume $\tau_{\lambda}\propto\lambda^{-0.7}$ we can compare a model of \citet{macarthur05} to our sample in Figure \ref{fig:dfn_irx}. This is shown as a dashed line in Figure \ref{fig:dfn_irx}. This same trend can be seen in the data in Figure \ref{fig:dfn_color} where the envelope of low \dfn~moves to larger values as the color (i.e. dust attenuation) increases. This shows the degree of reddening of \dfn~for large values of attenuation, and gives the deviation of the reddening vector from an exactly horizontal line in this diagram.

\begin{figure*}[t]
\epsscale{1.0}
\plotone{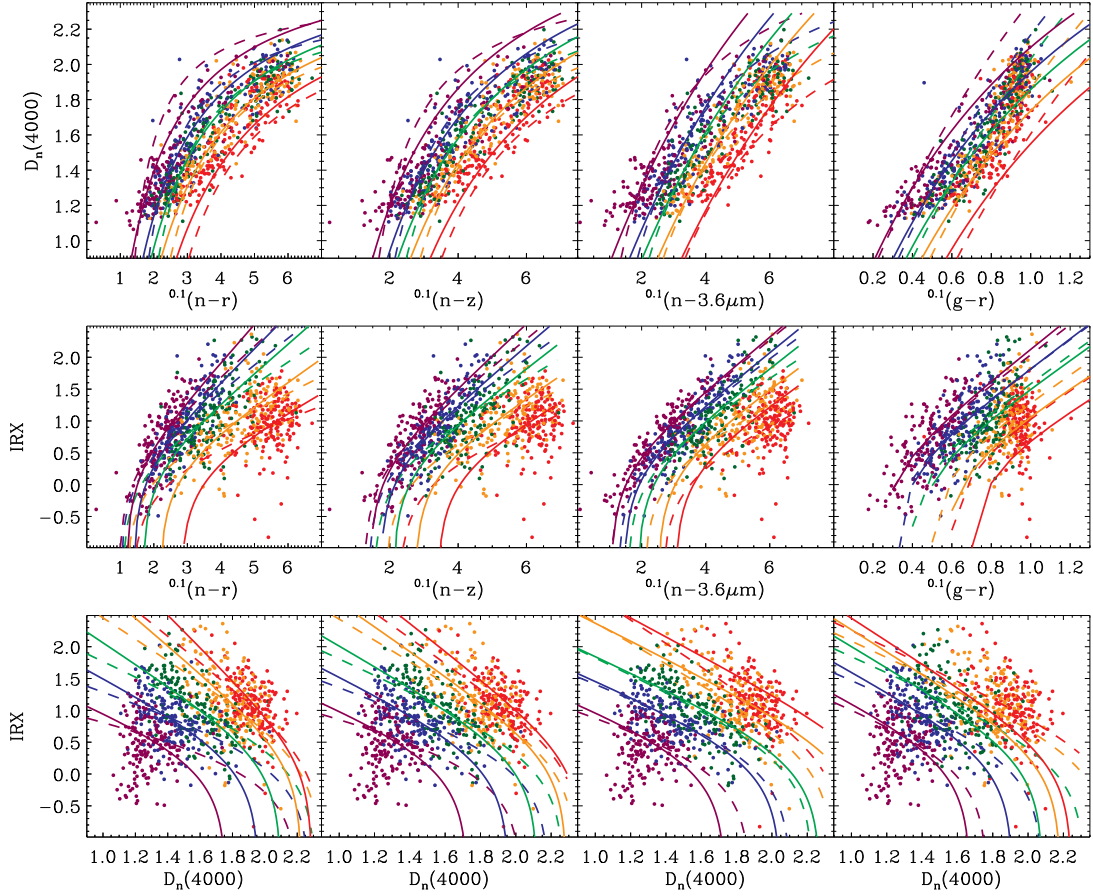}
\figcaption{Fits of Table \ref{tbl:fits} overplotted on the 3 projections of the relation between IRX, \dfn, and color.  From left to right the colors used are $^{0.1}(n-r)$, $^{0.1}(n-z)$, $^{0.1}(n-3.6\micron)$, $^{0.1}(g-r)$.    The solid lines show the fits determined from galaxies with \dfn$<1.6$ and $^{0.1}(n-r)<4$, while the dashed lines show the fits determined from the entire sample.
\label{fig:fit}}
\end{figure*}

\subsection{ Parameterization of the Relation\label{sec:fit}}

The relation between attenuation, color, and SFH described above is determined by the dust attenuation law and the correlation of \dfn~with unreddened color. We leave a more rigorous derivation of a dust law to a separate work \citep{paper3}. Here we give a simple parameterization of the relation between IRX, color, and \dfn~for a variety of colors. We fit using \airx~as the measure of dust attenuation since this is linear in color, and should closely approximate the true $A_{fuv}$ for relatively blue galaxies (\S\ref{sec:irx}). The relation between \dfn~and color is allowed to be linear or quadratic, based on visual inspection of Figure \ref{fig:dfn_color}. We are here assuming that the relation between \dfn~and the unreddened color is the same for all galaxies. Finally, after inspection of the fit residuals, we allow a cross term whose origin is unclear, but may be related to the change in the relation between $A_{fuv}$ and \airx~ as a function of SFH. Since the errors in \airx~ are larger than those in color or \dfn, we treat \airx~ as the dependent variable.  Thus, we fit

\begin{equation}
\label{eqn:fit}
\widehat{A}_{IRX}=\mbox{A}+\mbox{B}x+\mbox{C}y+\mbox{D}x^{2}+\mbox{E}xy
\end{equation}
where $x=$\dfn$-1.25$ and $y$ is the AB color, minus 2. The A term thus provides the typical attenuation for galaxies in the sample with \dfn$=1.25$ and a color of $\sim2$. The results of these fits are given in Table \ref{tbl:fits} for several different colors. We have also conducted fits after restricting the sample to \dfn$< 1.6$ and $^{0.1}(n-r)<4$. This allows us to ignore the effect of `older' galaxies for which \airx~ is likely not a good indicator of attenuation, and for which dust reddening may play a smaller role in determining the color, than, e.g. metallicity, AGN, or blue evolved stars. The inclusion of these high \dfn~galaxies may bias the fits, and when considering only blue sequence galaxies we encourage the reader to use the fits restricted to \dfn$< 1.6$.

\begin{figure}[h]
\epsscale{1.0}
\plotone{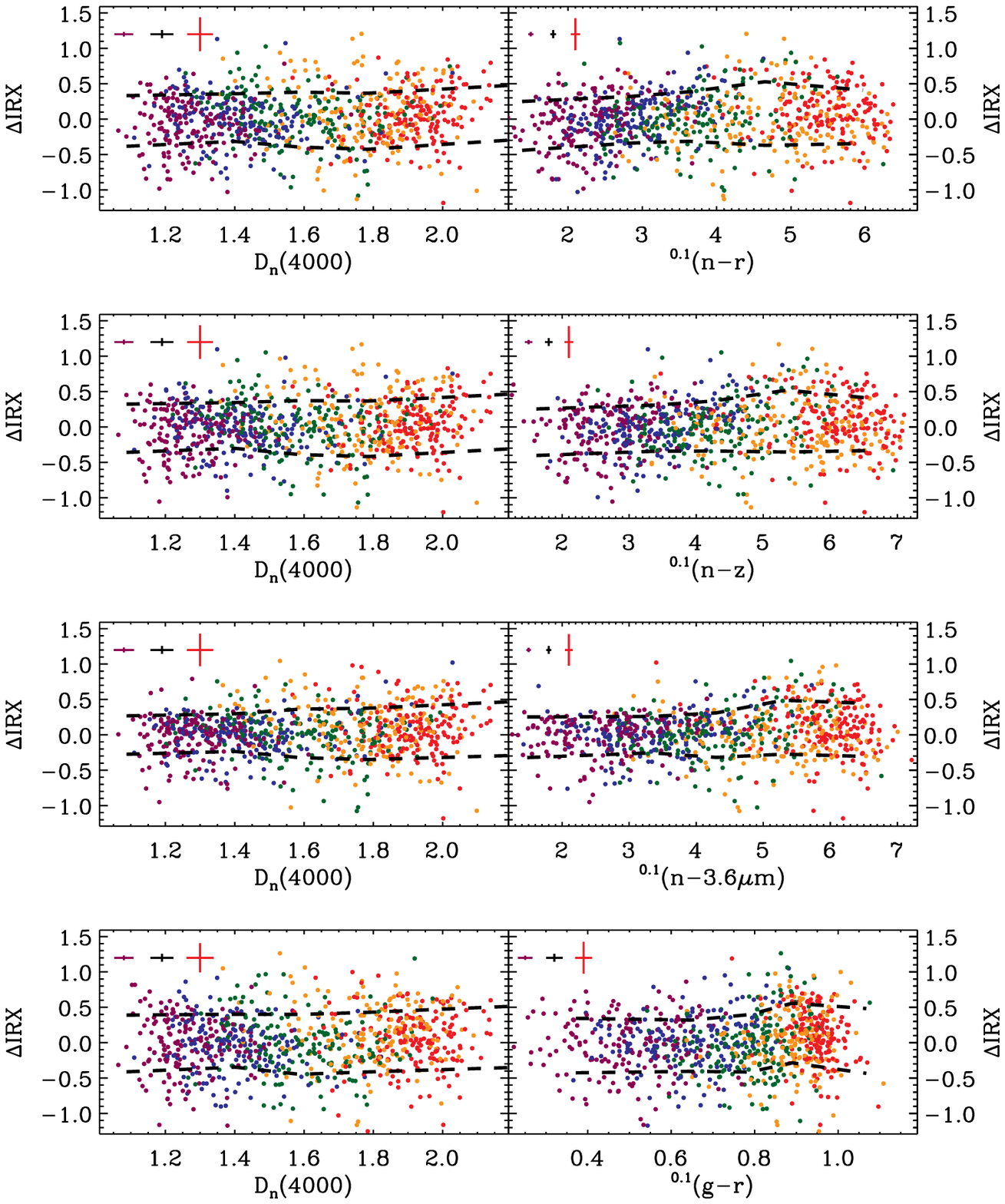}
\figcaption{Residuals from the fit (Table \ref{tbl:fits}) for each color, plotted vs \dfn~and color.  The residuals shown are from the fit to the entire sample including the A, B, C, and E terms in Table \ref{tbl:fits}.  Dashed lines show the standard deviation of the residuals.  The colors of the symbols in the left panels give the quintile of the observed color, while the color of symbols in the right panels gives the \dfn~quintile (see Figure \ref{fig:irx_color}, bottom right panel).
\label{fig:fit_resid}}
\end{figure}

\begin{figure*}[ht]
\epsscale{0.8}
\plotone{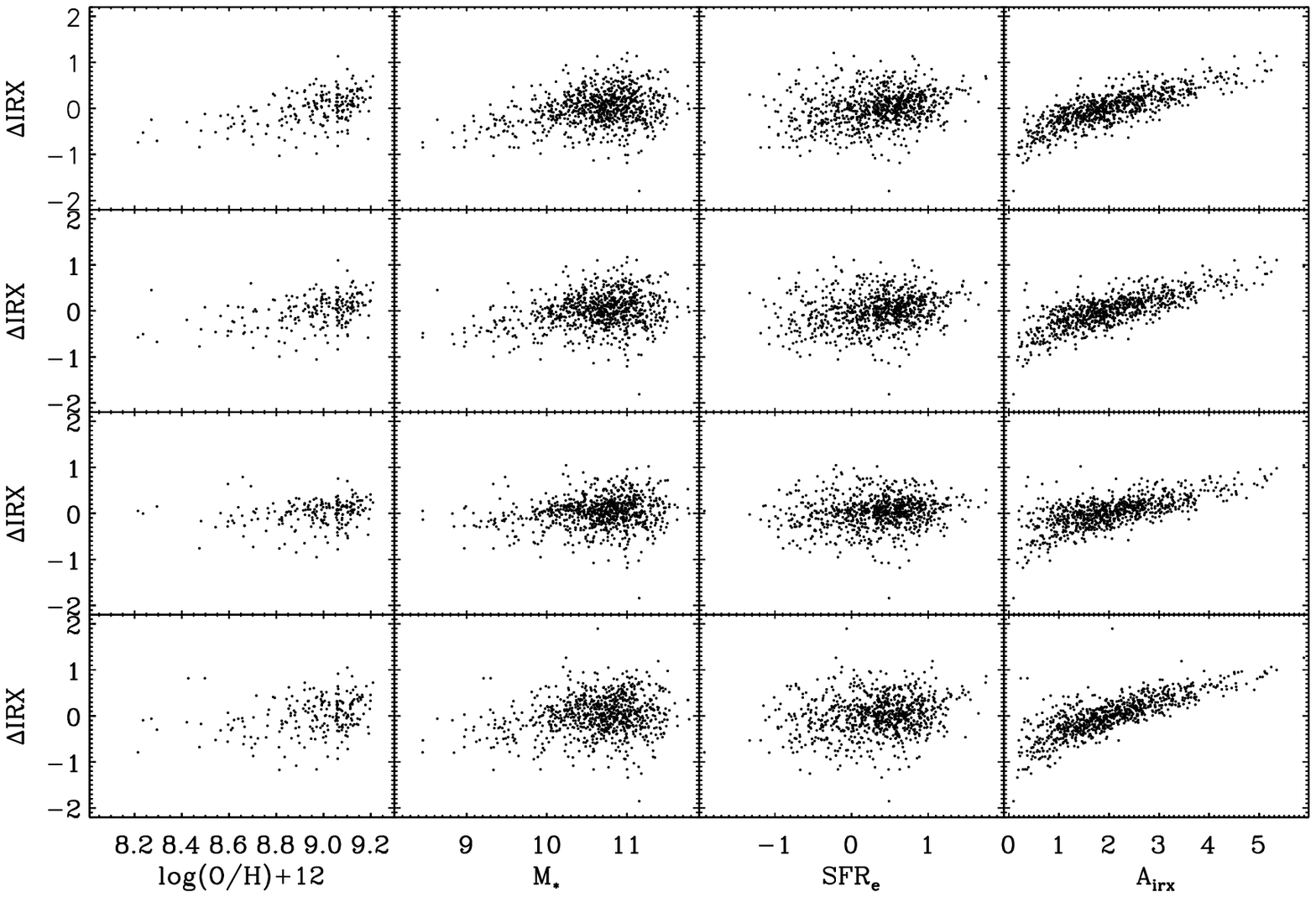}
\figcaption{ Residuals from the fit to the full sample, including the A, B, C, and E terms in Table \ref{tbl:fits}, plotted against other galaxy parameters. The colors used in the fit are ({\it top to bottom}), $^{0.1}(n-r)$, $^{0.1}(n-z)$, $^{0.1}(n-3.6\micron)$, and $^{0.1}(g-r)$. The residuals are plotted against ({\it left to right}), the gas-phase metallicity (where available), the stellar mass, the optically determined SFR \protect\citep[from][]{brinchmann04}, and \airx.
\label{fig:fit_resid2}}
\end{figure*}

The fits including a cross term (E) but without a quadratic term in \dfn~(D$=0$) are shown in Figure \ref{fig:fit} along with the sample galaxies.  This figure shows fits for both the entire sample and for the \dfn$<1.6$ subsample, in all three projections of the relation, and for the four colors given in Table \ref{tbl:fits}.  Because the displayed fits do not include D term, the curvature seen in all colors is due to the cross-term, although for $^{0.1}(n-3.6\micron)$ the fit to galaxies with \dfn$<1.6$ appears very linear. The cross-term also causes the lines for different \airx~values to not be parallel. Despite the comparable dispersion in the residuals, the fits using other combinations of terms are not as well behaved.  Residuals of the fits to the entire sample of galaxies are shown as a function of \dfn~and color in Figure \ref{fig:fit_resid}.  It is possible that additional parameters play a role in the relation. For example, the metallicity of galaxies may have a significant additional effect on the color beyond the effects of SFH and attenuation.  As a preliminary investigation of this possibility we show the residuals from the fit using $^{0.1}(n-3.6\micron)$ plotted against other galaxy parameters in Figure \ref{fig:fit_resid2}.  These parameters are the gas-phase metallicity as determined by \citet{tremonti04}, the stellar mass from \citet{kauffmann03a}, the SFR as determined by \citet{brinchmann04} (where we use the median of the probability distribution function for the global SFR), and \airx.The strongest trend is with \airx, the very quantity being fit.  This is a result of the dispersion in the relation being a large fraction of the range, and precludes detailed analysis of the residuals. 

\subsection{Applications and Implications for Future Observations and Models}

The \airx-color-\dfn~relation may be used to determine the attenuation given \dfn~and a color.  Such derived attenuations may then be used to correct SFR indicators at other wavelengths.  While \dfn~is already a coarse measure of the star-formation activity of galaxies, it is sensitive to older stars than the $f$ band or \ha~luminosity.  The accuracy of the attenuation correction derived from the fits and applied to UV or \ha~luminosities then depends on whether the most recently formed stars are affected by the same dust as those stars contributing to the long wavelength separation color (as well as the accuracy of the fit for the chosen color, and whether the form of the relation between \dfn~and color varies with changes in the details of the SFH). \citet{treyer07} have compared the UV fluxes corrected for attenuation using these fits to the \ha~derived SFR.  They find good agreement, although there is a trend of the differences with the SFR. A direct comparison of the UV$+$IR derived SFR to the emission-line derived SFR is given in a separate work \citep{paper2}.

Here we compare the accuracy of this method for attenuation correction to methods based on emission lines or the UV color.  The standard deviation of the residuals from the fits, in terms of both \airx~and IRX, are given in Table \ref{tbl:fits}.  For the $^{0.1}(n-3.6\micron)$ color these are 0.5 in \airx~and 0.3 in IRX.  Attenuation may also be estimated from the Balmer decrement A$_{Balmer}=2.5\log \frac{\mbox{L}_{H\alpha}/\mbox{L}_{H\beta}}{2.88}$, where L$_{H\alpha}$ and L$_{H\beta}$ are the \ha~and hb~line luminosities, repectively \citep{kennicutt98, paper2}.  
The ratio \airx/A$_{Balmer}$ is $\sim 3-5$ \citep{paper2}.  Thus, to obtain a similar error in \airx~from the Balmer decrement requires S/N $\gtrsim 10$ in both the \ha~and \hb~emission lines (or an equivalent combination of S/N in the two lines).  Another common method of attenuation correction is provided by the A$_{fuv}$-$\beta$ relation \citep[e.g.,][]{MHC}, where $\beta$ is the UV spectral slope (similar to the $f-n$ color).  \citet{seibert05a} quote a scatter of $\sigma($A$_{fuv})=0.89$ in this relation when a range of galaxy types is considered.  One must also consider the large slope in the relation between $\beta$ and A$_{fuv}$.  An uncertainty of 0.1 in $f-n$ color (due to photometric errors or uncertainty in the $K$-correction) leads to an uncertainty of $\sim 0.8$ in A$_{fuv}$, even in the absence of scatter in the A$_{fuv}$-$\beta$ relation.  Such uncertainties in color (and \dfn) are much less important for the \airx-color-\dfn~relation, especially when UV-optical colors are used that have large wavelength separation and relatively small errors. 

Besides correcting SFR measures at various wavelengths for dust attenuation, it may be interesting to consider the distribution of dust attenuations themselves -- derived from the \airx-color-\dfn~relation -- as a probe of the star formation in galaxies. The attenuation may be expected to trace the product of metallicity and gas surface density\citep{bell03_radio, paper3}, which are both key diagnostics of the evolution of galaxies \citep[e.g.,][]{martin07}.  The fit presented in \S\ref{sec:fit} can be used to estimate the dust attenuation for large samples of galaxies, at a range of redshifts, where \dfn~and a broadband color are known - high S/N emission lines are not necessary.  


Several future observatories will add to the IR view of the universe provided by \emph{Spitzer}, including  \emph{AKARI} \citep[formerly \emph{Astro-F},][]{akari} and the proposed \emph{Wide-Field Infrared Survey Explorer} \citep[\emph{WISE};][]{wise}.  In particular, the flux limit of \emph{WISE} at 23\micron~($\sim 2.6$ mJy, 5$\sigma$) will enable the detection of $\sim20$\% of SDSS/MPA galaxies at this wavelength (i.e., $10^5$ galaxies), although these will be primarily dusty star-forming galaxies. The parameterization of the relation given above will provide a context for this much larger sample of galaxies that will be strongly affected by selection effects in the IR.  Combined with \emph{GALEX} imaging over the SDSS footprint, the large number of galaxies can be used to investigate deviations from the relation as a function of, e.g. AGN activity or morphology -- understanding such deviations will lead to improved modelling of attenuation and knowledge of the SFH. Such a large number of galaxies will also provide an unbiased view of the properties of rare types of galaxies (e.g., ULIRGS), which may or may not follow the \airx-color-\dfn~relation derived here for more common galaxies. LIRGs at $z=0.5$ will be detectable by \emph{WISE} over the entire sky at 23\micron, while the 12\micron~band of WISE will be able to probe the restframe 8\micron~PAH emission of these galaxies.  Furthermore, nearly all SDSS/MPA galaxies will be detected in the shortest wavelength band ($\sim 3.3$\micron). Such NIR photometry -- when combined with resframe near-UV photometry -- leads to the clearest relation between broadband color, \dfn, and IRX, making accurate determinations of IRX possible for $\sim 10^5$ galaxies.  
 
Finally, this parameterized relation between \dfn, IRX, and color will have to be reproduced by the next generation of galaxy models that seek to self-consistently treat dust absorption and emission in the context of SPS modeling \citep{sunrise, yuexing_art, grasil}.  The emprical relation given above provides an important diagnostic for such models since it links, in a relatively straightforward way, the stellar populations, the absorption of the light of these populations by dust, and the subsequent re-emission of that light by dust at IR wavelengths.


\section{Color-Color Relations at High Redshift}
\label{sec:highz}

At high redshift ($z\sim2$) several authors have attempted to distinguish between the effects of attenuation and SFH on the colors of galaxies using \emph{Spitzer} IRAC imaging of the restframe NIR \citep{wuyts06, kriek06}. These use a methodology similar to that of \citet{kauffmann03a} by comparing the locations of galaxies in color-color space to a locus of SPS models, and attributing differences to dust reddening.  \citet{wuyts06} have used the relatively reddening insensitive photometric break strength at 4000\AA~(sampling both \dfn~and the Balmer break) explicitly as one of the `colors', with the benefit that the direction of the reddening vector is then relatively well known.  The determination of IRX, as in the present work, gives empirical constraints on the length of the reddening vector, lending support to this method for SFH determination. 

\begin{figure}[h]
\epsscale{1.0}
\plotone{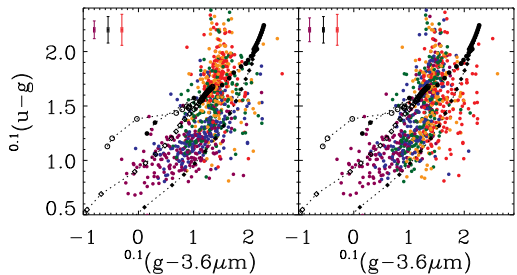}
\figcaption{Color-color diagrams as are available at high redshift using IRAC \citep{wuyts06}.  On the left the symbols are color-coded by the \dfn~quintile, on the right they are color coded by IRX quintile .  See Figures \ref{fig:irx_color} and \ref{fig:dfn_color} for the bins used for color coding, respectively.  The model symbols ({\it black}) are the same as in Figure \ref{fig:dfn_color}.  Colored error bars at the top left corner of each panel show the median errors for the corresponding parameter bin, while the black error bars show the median for the entire sample.
\label{fig:wuyts}}
\end{figure}

In Figure \ref{fig:wuyts} we show a color-color diagram for the sample galaxies that is quite similar to that constructed by \citet{wuyts06} for a number of galaxies at $z=1$-$3$, both in the colors plotted, and in the location of galaxies in the diagram.  Galaxies tend to move toward redder colors in $^{0.1}(g-3.6\micron)$ as IRX increases, and are generally at high \dfn~as $^{0.1}(u-g)$ increases.  As we have shown previously (\S\ref{sec:bc03}) there is also some reddening of the $u-g$ color with IRX. For a large portion of the color space (i.e.$^{0.1}(u-g)<1.5$), it is not possible use the color-color diagram alone to distinguish between different model tracks \emph{and} the amount of dust attenuation. With independent constraints on dust attenuation, and assuming that the population synthesis modeling yields the correct colors, it is possible to rule out high metallicity ($Z/Z_{\sun}=2.5$) exponentially declining models for a large proportion of the sample, although low metallicity exponentially declining models are consistent with the data.

\section{Conclusions}
\begin{enumerate}

\item We have presented a low redshift (\emph{\={z}}$=0.1$) context for the IR, UV, and optically derived dust attenuation, SF, and stellar mass properties of $\sim 1000$ homogeneously observed galaxies. The relation between SFH, attenuation, and color is a useful diagnostic of galaxy properties. We identified where special classes of galaxies lie in the ``color-color'' diagram of \dfn~versus broadband color.  We find that LIRGs occupy the same region of \dfn, IRX, and color parameter space as dusty star-forming galaxies, of which many are likely to be highly inclined.  We find that UVLGs occupy the same parameter space as a larger population of relatively attenuation free, star forming galaxies.  However, UVLGs are not the \emph{least} dusty galaxies for a given \dfn.  The distribution of galaxies in this parameter space, corrected for selection effects, can be compared to high redshift samples and semi-analytic models to determine the evolution of dust attenuation, and to insure that attenuation is being treated correctly in the models. 

\item This relation between \dfn, IRX, and color suggests that the colors of galaxies can be explained primarily as the contributions from the star formation history (we have here assumed exponentially declining models) and the dust attenuation \citep{dissecting}. A comparison with stellar population synthesis and dust attenuation models confirms that variation in \dfn~and IRX causes changes in galaxy color quantitatively similar to those seen in the data.  Detailed comparison of the exact form of the relation of \dfn~and IRX to the color of galaxies has the power to constrain these models. The UV color is especially sensistive to the form of the extinction law -- a Milky Way extinction law \citep{CCM} is not consistent with the change in UV colors of galaxies as a function of IRX, for a range of relative star/dust geometries \citep[but see][for a possible resolution involving the variation of attenuation with stellar age in galaxies]{panuzzo07}.

For some colors (e.g. $^{0.1}(u-r)$) the relation between \dfn~and color appears to be independent of the specific SFH model assumed.  Deviations of the observed galaxy colors from this relation may then be explained as reddening due to dust attenuation.  However, for other UV-optical-NIR colors (e.g. $^{0.1}(n-3.6\micron)$) the relation between \dfn~and color depends on the specific form of the SFH, especially at moderate \dfn~and color.  In this case, independent constraints on the degree of reddening provided by IRX will allow different SFH to be distinguished.  Similarly, different SFH models follow extremely different tracks in the \dfn~versus UV color plane, emphasizing the power of the UV color as a diagnostic of stellar populations.

\item We find through comparison with zero-dust stellar population synthesis models that a burst-like SFH leads to a relation between \dfn~and UV color at late times that is similar to that seen for our sample galaxies at \dfn$>1.7$. The colors of these galaxies do not appear to be highly correlated with IRX, suggesting that dust attenuation is not the dominant driver of UV color for these galaxies. This effect is likely caused by evolved blue stars. There is, however a significant offset of the locus of sample galaxies from the \citet{BC03} models of old stellar populations. 

\item We have parameterized the relation between \dfn, color, and IRX for a variety of colors.  This allows the determination of one of these quantities provided the other two are known. Assuming a universal relation between \dfn~and unattenuatted color (as provided by most exponentially declining SF models), this parametrization may be used to infer the distribution of IRX in galaxies, and the evolution thereof from high to low redshift.  It may also be used to determine the dust attenuations for application to tracers of more recent star formation.  The best accuracy in inferred attenuations is achieved with the $^{0.1}(n-3.6\micron)$ color -- \emph{AKARI} and the proposed \emph{WISE} observatory, combined with \emph{GALEX} will provide large samples of galaxies for which this color is available. This relation links dust absorption and emission and the stellar populations of galaxies.   Models that seek to explain the optical \emph{and} infrared emission of galaxies self-consistently will need to reproduce this relation for low-redshift galaxies. 

\end{enumerate}

\acknowledgments
The anonymous referee provided extremely useful comments that resulted in substantial improvements to the paper. BDJ thanks S. Salim, A. Boselli, S. Boissier, and L. Cortese for helpful comments.  BDJ was supported by NASA GSRP Grant NNG-05GO43H

\emph{GALEX} (\emph{Galaxy Evolution Explorer}) is a NASA Small Explorer, launched in 2003 April. We gratefully acknowledge NASA's support for construction, operation, and science analysis for the \emph{GALEX} mission, developed in cooperation with the Centre National d'Etudes Spatiales of France and the Korean Ministry of Science and Technology. 

This work is based in part on observations made with the Spitzer Space Telescope, which is operated by the Jet Propulsion Laboratory, California Institute of Technology under a contract with NASA.  In particular, the publicly available \emph{Spitzer} data obtained by the SWIRE team have been essential to this work.

Funding for the SDSS and SDSS-II has been provided by the Alfred P. Sloan Foundation, the Participating Institutions, the National Science Foundation, the U.S. Department of Energy, the National Aeronautics and Space Administration, the Japanese Monbukagakusho, the Max Planck Society, and the Higher Education Funding Council for England. The SDSS Web Site is http://www.sdss.org/. The SDSS is managed by the Astrophysical Research Consortium for the Participating Institutions. The Participating Institutions are the American Museum of Natural History, Astrophysical Institute Potsdam, University of Basel, University of Cambridge, Case Western Reserve University, University of Chicago, Drexel University, Fermilab, the Institute for Advanced Study, the Japan Participation Group, Johns Hopkins University, the Joint Institute for Nuclear Astrophysics, the Kavli Institute for Particle Astrophysics and Cosmology, the Korean Scientist Group, the Chinese Academy of Sciences (LAMOST), Los Alamos National Laboratory, the Max-Planck-Institute for Astronomy (MPIA), the Max-Planck-Institute for Astrophysics (MPA), New Mexico State University, Ohio State University, University of Pittsburgh, University of Portsmouth, Princeton University, the United States Naval Observatory, and the University of Washington.



{\it Facilities:} 

\bibliographystyle{apj}
\bibliography{ms}

\begin{thebibliography}{80}
\expandafter\ifx\csname natexlab\endcsname\relax\def\natexlab#1{#1}\fi

\bibitem[{{Abazajian} {et~al.}(2004){Abazajian}, {Adelman-McCarthy},
  {Ag{\"u}eros}, {Allam}, {Anderson}, {Anderson}, {Annis}, {Bahcall}, {Baldry},
  {Bastian}, {Berlind}, {Bernardi}, {Blanton}, {Bochanski}, {Boroski},
  {Briggs}, {Brinkmann}, {Brunner}, {Budav{\'a}ri}, {Carey}, {Carliles},
  {Castander}, {Connolly}, {Csabai}, {Doi}, {Dong}, {Eisenstein}, {Evans},
  {Fan}, {Finkbeiner}, {Friedman}, {Frieman}, {Fukugita}, {Gal}, {Gillespie},
  {Glazebrook}, {Gray}, {Grebel}, {Gunn}, {Gurbani}, {Hall}, {Hamabe},
  {Harris}, {Harris}, {Harvanek}, {Heckman}, {Hendry}, {Hennessy}, {Hindsley},
  {Hogan}, {Hogg}, {Holmgren}, {Ichikawa}, {Ichikawa}, {Ivezi{\'c}}, {Jester},
  {Johnston}, {Jorgensen}, {Kent}, {Kleinman}, {Knapp}, {Kniazev}, {Kron},
  {Krzesinski}, {Kunszt}, {Kuropatkin}, {Lamb}, {Lampeitl}, {Lee}, {Leger},
  {Li}, {Lin}, {Loh}, {Long}, {Loveday}, {Lupton}, {Malik}, {Margon},
  {Matsubara}, {McGehee}, {McKay}, {Meiksin}, {Munn}, {Nakajima}, {Nash},
  {Neilsen}, {Newberg}, {Newman}, {Nichol}, {Nicinski}, {Nieto-Santisteban},
  {Nitta}, {Okamura}, {O'Mullane}, {Ostriker}, {Owen}, {Padmanabhan},
  {Peoples}, {Pier}, {Pope}, {Quinn}, {Richards}, {Richmond}, {Rix}, {Rockosi},
  {Schlegel}, {Schneider}, {Scranton}, {Sekiguchi}, {Seljak}, {Sergey},
  {Sesar}, {Sheldon}, {Shimasaku}, {Siegmund}, {Silvestri}, {Smith}, {Smol{\v
  c}i{\'c}}, {Snedden}, {Stebbins}, {Stoughton}, {Strauss}, {SubbaRao},
  {Szalay}, {Szapudi}, {Szkody}, {Szokoly}, {Tegmark}, {Teodoro}, {Thakar},
  {Tremonti}, {Tucker}, {Uomoto}, {Vanden Berk}, {Vandenberg}, {Vogeley},
  {Voges}, {Vogt}, {Walkowicz}, {Wang}, {Weinberg}, {West}, {White}, {Wilhite},
  {Xu}, {Yanny}, {Yasuda}, {Yip}, {Yocum}, {York}, {Zehavi}, {Zibetti}, \&
  {Zucker}}]{sdss_dr2}
{Abazajian}, K. {et~al.} 2004, \aj, 128, 502

\bibitem[{{Baldry} {et~al.}(2004){Baldry}, {Glazebrook}, {Brinkmann},
  {Ivezi{\'c}}, {Lupton}, {Nichol}, \& {Szalay}}]{baldry}
{Baldry}, I.~K., {Glazebrook}, K., {Brinkmann}, J., {Ivezi{\'c}}, {\v Z}.,
  {Lupton}, R.~H., {Nichol}, R.~C., \& {Szalay}, A.~S. 2004, \apj, 600, 681

\bibitem[{{Balogh} {et~al.}(1999){Balogh}, {Morris}, {Yee}, {Carlberg}, \&
  {Ellingson}}]{balogh99}
{Balogh}, M.~L., {Morris}, S.~L., {Yee}, H.~K.~C., {Carlberg}, R.~G., \&
  {Ellingson}, E. 1999, \apj, 527, 54

\bibitem[{{Bell}(2002)}]{bell02b}
{Bell}, E.~F. 2002, \apj, 577, 150

\bibitem[{{Bell}(2003)}]{bell03_radio}
---. 2003, \apj, 586, 794

\bibitem[{{Bell} \& {de Jong}(2001)}]{bell_dejong}
{Bell}, E.~F., \& {de Jong}, R.~S. 2001, \apj, 550, 212

\bibitem[{{Bell} {et~al.}(2004{\natexlab{a}}){Bell}, {McIntosh}, {Barden},
  {Wolf}, {Caldwell}, {Rix}, {Beckwith}, {Borch}, {H{\"a}ussler}, {Jahnke},
  {Jogee}, {Meisenheimer}, {Peng}, {Sanchez}, {Somerville}, \&
  {Wisotzki}}]{bell_gemsdust}
{Bell}, E.~F. {et~al.} 2004{\natexlab{a}}, \apjl, 600, L11

\bibitem[{{Bell} {et~al.}(2004{\natexlab{b}}){Bell}, {Wolf}, {Meisenheimer},
  {Rix}, {Borch}, {Dye}, {Kleinheinrich}, {Wisotzki}, \&
  {McIntosh}}]{bell04_red}
---. 2004{\natexlab{b}}, \apj, 608, 752

\bibitem[{{Bertin} \& {Arnouts}(1996)}]{sextractor}
{Bertin}, E., \& {Arnouts}, S. 1996, \aaps, 117, 393

\bibitem[{{Blanton} {et~al.}(2003){Blanton}, {Hogg}, {Bahcall}, {Baldry},
  {Brinkmann}, {Csabai}, {Eisenstein}, {Fukugita}, {Gunn}, {Ivezi{\'c}},
  {Lamb}, {Lupton}, {Loveday}, {Munn}, {Nichol}, {Okamura}, {Schlegel},
  {Shimasaku}, {Strauss}, {Vogeley}, \& {Weinberg}}]{blanton03_prop}
{Blanton}, M.~R. {et~al.} 2003, \apj, 594, 186

\bibitem[{{Blanton} \& {Roweis}(2007)}]{blanton_k}
{Blanton}, M.~R., \& {Roweis}, S. 2007, \aj, 133, 734

\bibitem[{{Brinchmann} {et~al.}(2004){Brinchmann}, {Charlot}, {White},
  {Tremonti}, {Kauffmann}, {Heckman}, \& {Brinkmann}}]{brinchmann04}
{Brinchmann}, J., {Charlot}, S., {White}, S.~D.~M., {Tremonti}, C.,
  {Kauffmann}, G., {Heckman}, T., \& {Brinkmann}, J. 2004, \mnras, 351, 1151

\bibitem[{{Bruzual} \& {Charlot}(2003)}]{BC03}
{Bruzual}, G., \& {Charlot}, S. 2003, \mnras, 344, 1000

\bibitem[{{Buat}(1992)}]{buat92}
{Buat}, V. 1992, \aap, 264, 444

\bibitem[{{Buat} {et~al.}(2005){Buat}, {Iglesias-P{\'a}ramo}, {Seibert},
  {Burgarella}, {Charlot}, {Martin}, {Xu}, {Heckman}, {Boissier}, {Boselli},
  {Barlow}, {Bianchi}, {Byun}, {Donas}, {Forster}, {Friedman}, {Jelinski},
  {Lee}, {Madore}, {Malina}, {Milliard}, {Morissey}, {Neff}, {Rich},
  {Schiminovitch}, {Siegmund}, {Small}, {Szalay}, {Welsh}, \& {Wyder}}]{buat05}
{Buat}, V. {et~al.} 2005, \apjl, 619, L51

\bibitem[{{Buat} {et~al.}(2007){Buat}, {Takeuchi}, {Iglesias-Paramo}, {Xu},
  {Burgarella}, {Boselli}, {Barlow}, {Bianchi}, {Donas}, {Forster}, {Friedman},
  {Heckman}, {Lee}, {Madore}, {Martin}, {Milliard}, {Morissey}, {Neff}, {Rich},
  {Schiminovich}, {Seibert}, {Small}, {Szalay}, {Welsh}, {Wyder}, \&
  {Yi}}]{buat07}
---. 2007, \apjs, 173, 404

\bibitem[{{Buat} \& {Xu}(1996)}]{buat96}
{Buat}, V., \& {Xu}, C. 1996, \aap, 306, 61

\bibitem[{{Calzetti} {et~al.}(1994){Calzetti}, {Kinney}, \&
  {Storchi-Bergmann}}]{calzetti94}
{Calzetti}, D., {Kinney}, A.~L., \& {Storchi-Bergmann}, T. 1994, \apj, 429, 582

\bibitem[{{Cardelli} {et~al.}(1989){Cardelli}, {Clayton}, \& {Mathis}}]{CCM}
{Cardelli}, J.~A., {Clayton}, G.~C., \& {Mathis}, J.~S. 1989, \apj, 345, 245

\bibitem[{{Chabrier}(2003)}]{chabrier}
{Chabrier}, G. 2003, \pasp, 115, 763

\bibitem[{{Charlot} \& {Fall}(2000)}]{charlot00}
{Charlot}, S., \& {Fall}, S.~M. 2000, \apj, 539, 718

\bibitem[{{Dale} {et~al.}(2005){Dale}, {Bendo}, {Engelbracht}, {Gordon},
  {Regan}, {Armus}, {Cannon}, {Calzetti}, {Draine}, {Helou}, {Joseph},
  {Kennicutt}, {Li}, {Murphy}, {Roussel}, {Walter}, {Hanson}, {Hollenbach},
  {Jarrett}, {Kewley}, {Lamanna}, {Leitherer}, {Meyer}, {Rieke}, {Rieke},
  {Sheth}, {Smith}, \& {Thornley}}]{dale05}
{Dale}, D.~A. {et~al.} 2005, \apj, 633, 857

\bibitem[{{Dale} {et~al.}(2006){Dale}, {Bendo}, {Engelbracht}, {Gordon},
  {Regan}, {Armus}, {Cannon}, {Calzetti}, {Draine}, {Helou}, {Joseph},
  {Kennicutt}, {Li}, {Murphy}, {Roussel}, {Walter}, {Hanson}, {Hollenbach},
  {Jarrett}, {Kewley}, {Lamanna}, {Leitherer}, {Meyer}, {Rieke}, {Rieke},
  {Sheth}, {Smith}, \& {Thornley}}]{dalex}
---. 2006, \apj, 633, 857

\bibitem[{{Dale} {et~al.}(2001){Dale}, {Helou}, {Contursi}, {Silbermann}, \&
  {Kolhatkar}}]{dale01}
{Dale}, D.~A., {Helou}, G., {Contursi}, A., {Silbermann}, N.~A., \&
  {Kolhatkar}, S. 2001, \apj, 549, 215

\bibitem[{{Devriendt} {et~al.}(1999){Devriendt}, {Guiderdoni}, \&
  {Sadat}}]{devriendt}
{Devriendt}, J.~E.~G., {Guiderdoni}, B., \& {Sadat}, R. 1999, \aap, 350, 381

\bibitem[{{Draine} \& {Li}(2007)}]{draine_li06}
{Draine}, B.~T., \& {Li}, A. 2007, \apj, 657, 810

\bibitem[{Faber {et~al.}(2005)}]{faber05}
Faber, S.~M., {et~al.} 2005, {astro-ph/0506044}

\bibitem[{{Fazio} {et~al.}(2004){Fazio}, {Hora}, {Allen}, {Ashby}, {Barmby},
  {Deutsch}, {Huang}, {Kleiner}, {Marengo}, {Megeath}, {Melnick}, {Pahre},
  {Patten}, {Polizotti}, {Smith}, {Taylor}, {Wang}, {Willner}, {Hoffmann},
  {Pipher}, {Forrest}, {McMurty}, {McCreight}, {McKelvey}, {McMurray}, {Koch},
  {Moseley}, {Arendt}, {Mentzell}, {Marx}, {Losch}, {Mayman}, {Eichhorn},
  {Krebs}, {Jhabvala}, {Gezari}, {Fixsen}, {Flores}, {Shakoorzadeh}, {Jungo},
  {Hakun}, {Workman}, {Karpati}, {Kichak}, {Whitley}, {Mann}, {Tollestrup},
  {Eisenhardt}, {Stern}, {Gorjian}, {Bhattacharya}, {Carey}, {Nelson},
  {Glaccum}, {Lacy}, {Lowrance}, {Laine}, {Reach}, {Stauffer}, {Surace},
  {Wilson}, {Wright}, {Hoffman}, {Domingo}, \& {Cohen}}]{irac}
{Fazio}, G.~G. {et~al.} 2004, \apjs, 154, 10

\bibitem[{{Gil de Paz} {et~al.}(2007){Gil de Paz}, {Boissier}, {Madore},
  {Seibert}, {Joe}, {Boselli}, {Wyder}, {Thilker}, {Bianchi}, {Rey}, {Rich},
  {Barlow}, {Conrow}, {Forster}, {Friedman}, {Martin}, {Morrissey}, {Neff},
  {Schiminovich}, {Small}, {Donas}, {Heckman}, {Lee}, {Milliard}, {Szalay}, \&
  {Yi}}]{gil_de_paz}
{Gil de Paz}, A. {et~al.} 2007, \apjs, 173, 185

\bibitem[{{Gordon} {et~al.}(2000){Gordon}, {Clayton}, {Witt}, \&
  {Misselt}}]{gordon00}
{Gordon}, K.~D., {Clayton}, G.~C., {Witt}, A.~N., \& {Misselt}, K.~A. 2000,
  \apj, 533, 236

\bibitem[{{Gordon} {et~al.}(2007){Gordon}, {Engelbracht}, {Fadda},
  {Stansberry}, {Wachter}, {Frayer}, {Rieke}, {Noriega-Crespo}, {Latter},
  {Young}, {Neugebauer}, {Balog}, {Beeman}, {Dole}, {Egami}, {Haller}, {Hines},
  {Kelly}, {Marleau}, {Misselt}, {Morrison}, {P{\'e}rez-Gonz{\'a}lez}, {Rho},
  \& {Wheaton}}]{mips70}
{Gordon}, K.~D. {et~al.} 2007, \pasp, 119, 1019

\bibitem[{{Heckman} {et~al.}(2005){Heckman}, {Hoopes}, {Seibert}, {Martin},
  {Salim}, {Rich}, {Kauffmann}, {Charlot}, {Barlow}, {Bianchi}, {Byun},
  {Donas}, {Forster}, {Friedman}, {Jelinsky}, {Lee}, {Madore}, {Malina},
  {Milliard}, {Morrissey}, {Neff}, {Schiminovich}, {Siegmund}, {Small},
  {Szalay}, {Welsh}, \& {Wyder}}]{heckman05_uvlg}
{Heckman}, T.~M. {et~al.} 2005, \apjl, 619, L35

\bibitem[{{Helou} {et~al.}(2004){Helou}, {Roussel}, {Appleton}, {Frayer},
  {Stolovy}, {Storrie-Lombardi}, {Hurt}, {Lowrance}, {Makovoz}, {Masci},
  {Surace}, {Gordon}, {Alonso-Herrero}, {Engelbracht}, {Misselt}, {Rieke},
  {Rieke}, {Willner}, {Pahre}, {Ashby}, {Fazio}, \& {Smith}}]{helou04}
{Helou}, G. {et~al.} 2004, \apjs, 154, 253

\bibitem[{{Hirashita} {et~al.}(2003){Hirashita}, {Buat}, \&
  {Inoue}}]{hirashita03}
{Hirashita}, H., {Buat}, V., \& {Inoue}, A.~K. 2003, \aap, 410, 83

\bibitem[{{Iglesias-P{\'a}ramo} {et~al.}(2006){Iglesias-P{\'a}ramo}, {Buat},
  {Takeuchi}, {Xu}, {Boissier}, {Boselli}, {Burgarella}, {Madore}, {Gil de
  Paz}, {Bianchi}, {Barlow}, {Byun}, {Donas}, {Forster}, {Friedman}, {Heckman},
  {Jelinski}, {Lee}, {Malina}, {Martin}, {Milliard}, {Morrissey}, {Neff},
  {Rich}, {Schiminovich}, {Seibert}, {Siegmund}, {Small}, {Szalay}, {Welsh}, \&
  {Wyder}}]{iglesias_paramo06}
{Iglesias-P{\'a}ramo}, J. {et~al.} 2006, \apjs, 164, 38

\bibitem[{{Johnson} {et~al.}(2008){Johnson}, {Schiminovich}, {Seibert}, \&
  {Treyer}}]{ir_sed}
{Johnson}, B.~D., {Schiminovich}, D., {Seibert}, M., \& {Treyer}, M.~A. 2008,
  in preparation, 000, 000

\bibitem[{{Johnson} {et~al.}(2006){Johnson}, {Schiminovich}, {Seibert},
  {Treyer}, {Charlot}, {Heckman}, {Martin}, {Salim}, {Kauffmann}, {Bianchi},
  {Donas}, {Friedman}, {Lee}, {Madore}, {Milliard}, {Morrissey}, {Neff},
  {Rich}, {Szalay}, {Forster}, {Barlow}, {Conrow}, {Small}, \&
  {Wyder}}]{dissecting}
{Johnson}, B.~D. {et~al.} 2006, \apjl, 644, L109

\bibitem[{{Johnson} {et~al.}(2007{\natexlab{a}}){Johnson}, {Schiminovich},
  {Seibert}, {Treyer}, {Heckman}, {Martin}, {Bianchi}, {Donas}, {Friedman},
  {Lee}, {Madore}, {Milliard}, {Morrissey}, {Neff}, {Rich}, {Szalay},
  {Forster}, {Barlow}, {Conrow}, {Small}, \& {Wyder}}]{paper2}
---. 2007{\natexlab{a}}, \apjs, 644, 000

\bibitem[{{Johnson} {et~al.}(2007{\natexlab{b}}){Johnson}, {Schiminovich},
  {Seibert}, {Treyer}, {Heckman}, {Martin}, {Bianchi}, {Donas}, {Friedman},
  {Lee}, {Madore}, {Milliard}, {Morrissey}, {Neff}, {Rich}, {Szalay},
  {Forster}, {Barlow}, {Conrow}, {Small}, \& {Wyder}}]{paper3}
---. 2007{\natexlab{b}}, \apjs, 644, 000

\bibitem[{{Jonsson}(2006)}]{sunrise}
{Jonsson}, P. 2006, \mnras, 372, 2

\bibitem[{{Kauffmann} {et~al.}(2007){Kauffmann}, {Heckman}, {Budavari},
  {Charlot}, {Hoopes}, {Martin}, {Seibert}, {Barlow}, {Bianchi}, {Conrow},
  {Donas}, {Forster}, {Friedman}, {Lee}, {Madore}, {Milliard}, {Morrissey},
  {Neff}, {Rich}, {Schiminovich}, {Small}, {Szalay}, {Wyder}, \&
  {Yi}}]{kauffmann06}
{Kauffmann}, G. {et~al.} 2007, \apjs, 173, 357

\bibitem[{{Kauffmann} {et~al.}(2003{\natexlab{a}}){Kauffmann}, {Heckman},
  {Tremonti}, {Brinchmann}, {Charlot}, {White}, {Ridgway}, {Brinkmann},
  {Fukugita}, {Hall}, {Ivezi{\'c}}, {Richards}, \& {Schneider}}]{kauffmann03b}
---. 2003{\natexlab{a}}, \mnras, 346, 1055

\bibitem[{{Kauffmann} {et~al.}(2003{\natexlab{b}}){Kauffmann}, {Heckman},
  {White}, {Charlot}, {Tremonti}, {Brinchmann}, {Bruzual}, {Peng}, {Seibert},
  {Bernardi}, {Blanton}, {Brinkmann}, {Castander}, {Cs{\'a}bai}, {Fukugita},
  {Ivezic}, {Munn}, {Nichol}, {Padmanabhan}, {Thakar}, {Weinberg}, \&
  {York}}]{kauffmann03a}
---. 2003{\natexlab{b}}, \mnras, 341, 33

\bibitem[{{Kennicutt}(1998)}]{kennicutt98}
{Kennicutt}, R.~C. 1998, \araa, 36, 189

\bibitem[{{Kennicutt} {et~al.}(2003){Kennicutt}, {Armus}, {Bendo}, {Calzetti},
  {Dale}, {Draine}, {Engelbracht}, {Gordon}, {Grauer}, {Helou}, {Hollenbach},
  {Jarrett}, {Kewley}, {Leitherer}, {Li}, {Malhotra}, {Regan}, {Rieke},
  {Rieke}, {Roussel}, {Smith}, {Thornley}, \& {Walter}}]{kennicutt_sings}
{Kennicutt}, Jr., R.~C. {et~al.} 2003, \pasp, 115, 928

\bibitem[{{Kong} {et~al.}(2004){Kong}, {Charlot}, {Brinchmann}, \&
  {Fall}}]{kong04}
{Kong}, X., {Charlot}, S., {Brinchmann}, J., \& {Fall}, S.~M. 2004, \mnras,
  349, 769

\bibitem[{{Kriek} {et~al.}(2006){Kriek}, {van Dokkum}, {Franx}, {F{\"o}rster
  Schreiber}, {Gawiser}, {Illingworth}, {Labb{\'e}}, {Marchesini}, {Quadri},
  {Rix}, {Rudnick}, {Toft}, {van der Werf}, \& {Wuyts}}]{kriek06}
{Kriek}, M. {et~al.} 2006, \apj, 645, 44

\bibitem[{{Labb{\'e}} {et~al.}(2007){Labb{\'e}}, {Franx}, {Rudnick},
  {Schreiber}, {van Dokkum}, {Moorwood}, {Rix}, {R{\"o}ttgering}, {Trujillo},
  \& {van der Werf}}]{labbe07}
{Labb{\'e}}, I. {et~al.} 2007, \apj, 665, 944

\bibitem[{{Li} {et~al.}(2007){Li}, {Hopkins}, {Hernquist}, {Finkbeiner}, {Cox},
  {Springel}, {Jiang}, {Fan}, \& {Yoshida}}]{yuexing_art}
{Li}, Y. {et~al.} 2007, ArXiv e-prints, 706

\bibitem[{{Lonsdale} {et~al.}(2003){Lonsdale}, {Smith}, {Rowan-Robinson},
  {Surace}, {Shupe}, {Xu}, {Oliver}, {Padgett}, {Fang}, {Conrow},
  {Franceschini}, {Gautier}, {Griffin}, {Hacking}, {Masci}, {Morrison},
  {O'Linger}, {Owen}, {P{\'e}rez-Fournon}, {Pierre}, {Puetter}, {Stacey},
  {Castro}, {Del Carmen Polletta}, {Farrah}, {Jarrett}, {Frayer}, {Siana},
  {Babbedge}, {Dye}, {Fox}, {Gonzalez-Solares}, {Salaman}, {Berta}, {Condon},
  {Dole}, \& {Serjeant}}]{swire}
{Lonsdale}, C.~J. {et~al.} 2003, \pasp, 115, 897

\bibitem[{{MacArthur}(2005)}]{macarthur05}
{MacArthur}, L.~A. 2005, \apj, 623, 795

\bibitem[{{Mainzer} {et~al.}(2005){Mainzer}, {Eisenhardt}, {Wright}, {Liu},
  {Irace}, {Heinrichsen}, {Cutri}, \& {Duval}}]{wise}
{Mainzer}, A.~K., {Eisenhardt}, P., {Wright}, E.~L., {Liu}, F.-C., {Irace}, W.,
  {Heinrichsen}, I., {Cutri}, R., \& {Duval}, V. 2005, in Presented at the
  Society of Photo-Optical Instrumentation Engineers (SPIE) Conference, Vol.
  5899, UV/Optical/IR Space Telescopes: Innovative Technologies and Concepts
  II. Edited by MacEwen, Howard A. Proceedings of the SPIE, Volume 5899, pp.
  262-273 (2005)., ed. H.~A. {MacEwen}, 262--273

\bibitem[{{Martin} {et~al.}(2006){Martin}, {Seibert}, {Buat},
  {Iglesias-P{\'a}ramo}, {Barlow}, {Bianchi}, {Byun}, {Donas}, {Forster},
  {Friedman}, {Heckman}, {Jelinsky}, {Lee}, {Madore}, {Malina}, {Milliard},
  {Morrissey}, {Neff}, {Rich}, {Schiminovich}, {Siegmund}, {Small}, {Szalay},
  {Welsh}, \& {Wyder}}]{martin07}
{Martin}, D.~C. {et~al.} 2006, \apjl, 619, L59

\bibitem[{{Meurer} {et~al.}(1999){Meurer}, {Heckman}, \& {Calzetti}}]{MHC}
{Meurer}, G.~R., {Heckman}, T.~M., \& {Calzetti}, D. 1999, \apj, 521, 64

\bibitem[{{Mobasher} {et~al.}(2004){Mobasher}, {Idzi}, {Ben{\'{\i}}tez},
  {Cimatti}, {Cristiani}, {Daddi}, {Dahlen}, {Dickinson}, {Erben}, {Ferguson},
  {Giavalisco}, {Grogin}, {Koekemoer}, {Mignoli}, {Moustakas}, {Nonino},
  {Rosati}, {Schirmer}, {Stern}, {Vanzella}, {Wolf}, \&
  {Zamorani}}]{mobasher_photoz}
{Mobasher}, B. {et~al.} 2004, \apjl, 600, L167

\bibitem[{{Morrissey} {et~al.}(2007){Morrissey}, {Conrow}, {Barlow}, {Small},
  {Seibert}, {Wyder}, {Budavari}, {Arnouts}, {Friedman}, {Forster}, {Martin},
  {Neff}, {Schiminovich}, {Bianchi}, {Donas}, {Heckman}, {Lee}, {Madore}, \&
  {Milliard}}]{galex_pipe}
{Morrissey}, P. {et~al.} 2007, \apjs, 173, 682

\bibitem[{{Moustakas} \& {Kennicutt}(2006)}]{moustakas}
{Moustakas}, J., \& {Kennicutt}, Jr., R.~C. 2006, \apjs, 164, 81

\bibitem[{{Murakami}(1998)}]{akari}
{Murakami}, H. 1998, in Presented at the Society of Photo-Optical
  Instrumentation Engineers (SPIE) Conference, Vol. 3356, Proc. SPIE Vol. 3356,
  p. 471-477, Space Telescopes and Instruments V, Pierre Y. Bely; James B.
  Breckinridge; Eds., ed. P.~Y. {Bely} \& J.~B. {Breckinridge}, 471--477

\bibitem[{{Noeske} {et~al.}(2007){Noeske}, {Faber}, {Weiner}, {Koo}, {Primack},
  {Dekel}, {Papovich}, {Conselice}, {Le Floc'h}, {Rieke}, {Coil}, {Lotz},
  {Somerville}, \& {Bundy}}]{noeske07b}
{Noeske}, K.~G. {et~al.} 2007, \apjl, 660, L47

\bibitem[{{Panuzzo} {et~al.}(2007){Panuzzo}, {Granato}, {Buat}, {Inoue},
  {Silva}, {Iglesias-P{\'a}ramo}, \& {Bressan}}]{panuzzo07}
{Panuzzo}, P., {Granato}, G.~L., {Buat}, V., {Inoue}, A.~K., {Silva}, L.,
  {Iglesias-P{\'a}ramo}, J., \& {Bressan}, A. 2007, \mnras, 375, 640

\bibitem[{{Papovich} \& {Bell}(2002)}]{papovich02}
{Papovich}, C., \& {Bell}, E.~F. 2002, \apjl, 579, L1

\bibitem[{{Pierini} {et~al.}(2004){Pierini}, {Gordon}, {Witt}, \&
  {Madsen}}]{pierini04}
{Pierini}, D., {Gordon}, K.~D., {Witt}, A.~N., \& {Madsen}, G.~J. 2004, \apj,
  617, 1022

\bibitem[{{Rieke} {et~al.}(2004){Rieke}, {Young}, {Engelbracht}, {Kelly},
  {Low}, {Haller}, {Beeman}, {Gordon}, {Stansberry}, {Misselt}, {Cadien},
  {Morrison}, {Rivlis}, {Latter}, {Noriega-Crespo}, {Padgett}, {Stapelfeldt},
  {Hines}, {Egami}, {Muzerolle}, {Alonso-Herrero}, {Blaylock}, {Dole}, {Hinz},
  {Le Floc'h}, {Papovich}, {P{\'e}rez-Gonz{\'a}lez}, {Smith}, {Su}, {Bennett},
  {Frayer}, {Henderson}, {Lu}, {Masci}, {Pesenson}, {Rebull}, {Rho}, {Keene},
  {Stolovy}, {Wachter}, {Wheaton}, {Werner}, \& {Richards}}]{mips}
{Rieke}, G.~H. {et~al.} 2004, \apjs, 154, 25

\bibitem[{{Salim} {et~al.}(2006){Salim}, {Charlot}, {Rich}, {Kauffmann},
  {Heckman}, {Barlow}, {Bianchi}, {Byun}, {Donas}, {Forster}, {Friedman},
  {Jelinsky}, {Lee}, {Madore}, {Malina}, {Martin}, {Milliard}, {Morrissey},
  {Neff}, {Schiminovich}, {Seibert}, {Siegmund}, {Small}, {Szalay}, {Welsh}, \&
  {Wyder}}]{salim07}
{Salim}, S. {et~al.} 2006, \apjs, 619, L39

\bibitem[{{Schiminovich} {et~al.}(2007){Schiminovich}, {Wyder}, {Martin},
  {Johnson}, {Salim}, {Seibert}, {Treyer}, {Budavari}, {Hoopes}, {Zamojski},
  {Barlow}, {Forster}, {Friedman}, {Morrissey}, {Neff}, {Small}, {Bianchi},
  {Donas}, \& {Heckman}}]{schiminovich07}
{Schiminovich}, D. {et~al.} 2007, \apjs, 173, 315

\bibitem[{{Schlegel} {et~al.}(1998){Schlegel}, {Finkbeiner}, \& {Davis}}]{SFD}
{Schlegel}, D.~J., {Finkbeiner}, D.~P., \& {Davis}, M. 1998, \apj, 500, 525

\bibitem[{{Schweizer} \& {Seitzer}(1992)}]{ss92}
{Schweizer}, F., \& {Seitzer}, P. 1992, \aj, 104, 1039

\bibitem[{{Seibert} {et~al.}(2005){Seibert}, {Martin}, {Heckman}, {Buat},
  {Hoopes}, {Barlow}, {Bianchi}, {Byun}, {Donas}, {Forster}, {Friedman},
  {Jelinsky}, {Lee}, {Madore}, {Malina}, {Milliard}, {Morrissey}, {Neff},
  {Rich}, {Schiminovich}, {Siegmund}, {Small}, {Szalay}, {Welsh}, \&
  {Wyder}}]{seibert05a}
{Seibert}, M. {et~al.} 2005, \apjl, 619, L55

\bibitem[{{Silva} {et~al.}(1998){Silva}, {Granato}, {Bressan}, \&
  {Danese}}]{grasil}
{Silva}, L., {Granato}, G.~L., {Bressan}, A., \& {Danese}, L. 1998, \apj, 509,
  103

\bibitem[{{Smith} {et~al.}(2007){Smith}, {Draine}, {Dale}, {Moustakas},
  {Kennicutt}, {Helou}, {Armus}, {Roussel}, {Sheth}, {Bendo}, {Buckalew},
  {Calzetti}, {Engelbracht}, {Gordon}, {Hollenbach}, {Li}, {Malhotra},
  {Murphy}, \& {Walter}}]{smith06}
{Smith}, J.~D.~T. {et~al.} 2007, \apj, 656, 770

\bibitem[{{Stecher}(1965)}]{dust_bump}
{Stecher}, T.~P. 1965, \apj, 142, 1683

\bibitem[{{Stoughton} {et~al.}(2002){Stoughton}, {Lupton}, {Bernardi},
  {Blanton}, {Burles}, {Castander}, {Connolly}, {Eisenstein}, {Frieman},
  {Hennessy}, {Hindsley}, {Ivezi{\'c}}, {Kent}, {Kunszt}, {Lee}, {Meiksin},
  {Munn}, {Newberg}, {Nichol}, {Nicinski}, {Pier}, {Richards}, {Richmond},
  {Schlegel}, {Smith}, {Strauss}, {SubbaRao}, {Szalay}, {Thakar}, {Tucker},
  {Vanden Berk}, {Yanny}, {Adelman}, {Anderson}, {Anderson}, {Annis},
  {Bahcall}, {Bakken}, {Bartelmann}, {Bastian}, {Bauer}, {Berman},
  {B{\"o}hringer}, {Boroski}, {Bracker}, {Briegel}, {Briggs}, {Brinkmann},
  {Brunner}, {Carey}, {Carr}, {Chen}, {Christian}, {Colestock}, {Crocker},
  {Csabai}, {Czarapata}, {Dalcanton}, {Davidsen}, {Davis}, {Dehnen},
  {Dodelson}, {Doi}, {Dombeck}, {Donahue}, {Ellman}, {Elms}, {Evans}, {Eyer},
  {Fan}, {Federwitz}, {Friedman}, {Fukugita}, {Gal}, {Gillespie}, {Glazebrook},
  {Gray}, {Grebel}, {Greenawalt}, {Greene}, {Gunn}, {de Haas}, {Haiman},
  {Haldeman}, {Hall}, {Hamabe}, {Hansen}, {Harris}, {Harris}, {Harvanek},
  {Hawley}, {Hayes}, {Heckman}, {Helmi}, {Henden}, {Hogan}, {Hogg}, {Holmgren},
  {Holtzman}, {Huang}, {Hull}, {Ichikawa}, {Ichikawa}, {Johnston}, {Kauffmann},
  {Kim}, {Kimball}, {Kinney}, {Klaene}, {Kleinman}, {Klypin}, {Knapp},
  {Korienek}, {Krolik}, {Kron}, {Krzesi{\'n}ski}, {Lamb}, {Leger},
  {Limmongkol}, {Lindenmeyer}, {Long}, {Loomis}, {Loveday}, {MacKinnon},
  {Mannery}, {Mantsch}, {Margon}, {McGehee}, {McKay}, {McLean}, {Menou},
  {Merelli}, {Mo}, {Monet}, {Nakamura}, {Narayanan}, {Nash}, {Neilsen},
  {Newman}, {Nitta}, {Odenkirchen}, {Okada}, {Okamura}, {Ostriker}, {Owen},
  {Pauls}, {Peoples}, {Peterson}, {Petravick}, {Pope}, {Pordes}, {Postman},
  {Prosapio}, {Quinn}, {Rechenmacher}, {Rivetta}, {Rix}, {Rockosi}, {Rosner},
  {Ruthmansdorfer}, {Sandford}, {Schneider}, {Scranton}, {Sekiguchi}, {Sergey},
  {Sheth}, {Shimasaku}, {Smee}, {Snedden}, {Stebbins}, {Stubbs}, {Szapudi},
  {Szkody}, {Szokoly}, {Tabachnik}, {Tsvetanov}, {Uomoto}, {Vogeley}, {Voges},
  {Waddell}, {Walterbos}, {Wang}, {Watanabe}, {Weinberg}, {White}, {White},
  {Wilhite}, {Wolfe}, {Yasuda}, {York}, {Zehavi}, \& {Zheng}}]{sdss_edr}
{Stoughton}, C. {et~al.} 2002, \aj, 123, 485

\bibitem[{{Strauss} {et~al.}(2002){Strauss}, {Weinberg}, {Lupton}, {Narayanan},
  {Annis}, {Bernardi}, {Blanton}, {Burles}, {Connolly}, {Dalcanton}, {Doi},
  {Eisenstein}, {Frieman}, {Fukugita}, {Gunn}, {Ivezi{\'c}}, {Kent}, {Kim},
  {Knapp}, {Kron}, {Munn}, {Newberg}, {Nichol}, {Okamura}, {Quinn}, {Richmond},
  {Schlegel}, {Shimasaku}, {SubbaRao}, {Szalay}, {Vanden Berk}, {Vogeley},
  {Yanny}, {Yasuda}, {York}, \& {Zehavi}}]{strauss02}
{Strauss}, M.~A. {et~al.} 2002, \aj, 124, 1810

\bibitem[{{Tinsley}(1968)}]{tinsley68}
{Tinsley}, B.~M. 1968, \apj, 151, 547

\bibitem[{{Tremonti} {et~al.}(2004){Tremonti}, {Heckman}, {Kauffmann},
  {Brinchmann}, {Charlot}, {White}, {Seibert}, {Peng}, {Schlegel}, {Uomoto},
  {Fukugita}, \& {Brinkmann}}]{tremonti04}
{Tremonti}, C.~A. {et~al.} 2004, \apj, 613, 898

\bibitem[{{Treyer} {et~al.}(2007){Treyer}, {Schiminovich}, {Johnson},
  {Seibert}, {Wyder}, {Barlow}, {Conrow}, {Forster}, {Friedman}, {Martin},
  {Morrissey}, {Neff}, {Small}, {Bianchi}, {Donas}, {Heckman}, {Lee}, {Madore},
  {Milliard}, {Rich}, {Szalay}, {Welsh}, \& {Yi}}]{treyer07}
{Treyer}, M. {et~al.} 2007, \apjs, 137, 256

\bibitem[{{Wang} \& {Heckman}(1996)}]{WH96}
{Wang}, B., \& {Heckman}, T.~M. 1996, \apj, 457, 645

\bibitem[{{Witt} \& {Gordon}(2000)}]{WG00}
{Witt}, A.~N., \& {Gordon}, K.~D. 2000, \apj, 528, 799

\bibitem[{{Wuyts} {et~al.}(2007){Wuyts}, {Labb{\'e}}, {Franx}, {Rudnick}, {van
  Dokkum}, {Fazio}, {F{\"o}rster Schreiber}, {Huang}, {Moorwood}, {Rix},
  {R{\"o}ttgering}, \& {van der Werf}}]{wuyts06}
{Wuyts}, S. {et~al.} 2007, \apj, 655, 51

\bibitem[{{Wyder} {et~al.}(2007){Wyder}, {Martin}, {Schiminovich}, {Seibert},
  {Budavari}, {Treyer}, {Barlow}, {Forster}, {Friedman}, {Morrissey}, {Neff},
  {Small}, {Bianchi}, {Donas}, {Heckman}, {Lee}, {Madore}, {Milliard}, {Rich},
  {Szalay}, {Welsh}, \& {Yi}}]{wyder07}
{Wyder}, T.~K. {et~al.} 2007, \apjs, 173, 293

\end{thebibliography}





\clearpage
\thispagestyle{empty}
\begin{deluxetable}{ccccccc|ccccccc}
\tablecolumns{14}
\tablewidth{0pc}
\tablecaption{Empirical Fits for Different Broadband Colors \label{tbl:fits}
}
\tablehead{}
\startdata
\hline
\hline
\multicolumn{7}{c}{All Galaxies} & \multicolumn{7}{c}{\dfn$<1.6$ and $^{0.1}(n-r)<4$ (401 galaxies)} \\
A & B & C & D & E & $\sigma$(\airx) & $\sigma$(IRX) & A & B & C & D & E & $\sigma$(\airx) & $\sigma$(IRX)\\
\cutinhead{$y=^{0.1}(n-r)-2$}
1.61 & -2.96 &  0.77 &  ---  &  ---  & 0.76 & 0.40 & 1.21 & -3.16 & 1.38 &  ---  &  ---  & 0.60 & 0.33 \\
1.48 & -1.12 &  0.76 & -2.50 &  ---  & 0.74 & 0.38 & 1.24 & -2.12 & 1.36 & -4.65 &  ---  & 0.57 & 0.32 \\
1.25 & -1.33 &  1.19 &  ---  & -1.02 & 0.70 & 0.35 & 1.20 & -2.51 & 1.48 &  ---  & -1.05 & 0.59 & 0.32 \\
1.20 & -3.35 &  1.57 & 4.69  & -1.91 & 0.68 & 0.35 & 1.22 & -2.23 & 1.44 & -2.00 & -0.79 & 0.59 & 0.32 \\
\cutinhead{$y=^{0.1}(n-z)-2$}
1.23 & -3.27 & 0.78 &  ---  &  ---  & 0.73 & 0.38 & 0.74 & -3.26 & 1.18 &  ---  &  ---  & 0.58 & 0.32 \\
1.13 & -1.66 & 0.76 & -2.14 &  ---  & 0.72 & 0.37 & 0.77 & -2.28 & 1.16 & -4.34 &  ---  & 0.57 & 0.32 \\
0.81 & -1.32 & 1.07 &  ---  & -0.81 & 0.68 & 0.35 & 0.71 & -2.49 & 1.24 &  ---  & -0.67 & 0.57 & 0.32 \\
0.65 & -2.79 & 1.36 &  4.22 & -1.53 & 0.66 & 0.35 & 0.75 & -2.20 & 1.20 & -2.90 & -0.35 & 0.57 & 0.32 \\
\cutinhead{$y=^{0.1}(n-3.6\micron)-2$}
1.25 & -3.49 & 0.84 &  ---  &  ---  & 0.63 & 0.35 & 0.98 & -2.92 & 1.03 &  ---  &  ---  & 0.46 & 0.28 \\
1.17 & -2.24 & 0.83 & -1.64 &  ---  & 0.62 & 0.32 & 1.00 & -2.28 & 1.02 & -2.87 &  ---  & 0.46 & 0.27 \\
0.98 & -1.89 & 1.02 &  ---  & -0.63 & 0.59 & 0.31 & 0.98 & -2.65 & 1.05 &  ---  & -0.25 & 0.46 & 0.27 \\
0.94 & -2.80 & 1.16 &  2.51 & -1.02 & 0.58 & 0.31 & 1.00 & -2.27 & 1.02 & -2.81 & -0.02 & 0.46 & 0.27 \\
\cutinhead{$y=^{0.1}(g-r)-2$}
10.23 & -3.17 &  5.75 &  ---  &  ---  & 0.74 & 0.38 & 10.60 & -3.71 & 5.99 &  ---  &  ---  & 0.67 & 0.36 \\
10.47 & -3.63 &  5.90 &  0.51 &  ---  & 0.74 & 0.38 & 10.46 & -2.85 & 5.89 & -3.67 &  ---  & 0.67 & 0.36 \\
10.35 & -4.34 &  5.85 &  ---  & -1.05 & 0.74 & 0.38 & 10.84 & -6.64 & 6.15 &  ---  & -2.10 & 0.67 & 0.36 \\
12.05 &-10.59 &  6.96 &  2.76 & -4.44 & 0.74 & 0.38 & 10.38 & -1.98 & 5.84 & -4.18 &  0.54 & 0.67 & 0.36 
\enddata
\end{deluxetable}







\end{document}